 \documentclass[12pt,preprint]{aastex}

\usepackage{amsmath}

\usepackage{paralist}

\usepackage{verbatim}

\usepackage{soul}
\usepackage[usenames,dvipsnames]{color}

\usepackage{amsbsy}

\usepackage{graphicx}

\shorttitle{Bayesian Approach to Kepler Cotrending}
\shortauthors{J. C. Smith et al.}

\begin{document}

\title{Kepler Presearch Data Conditioning II -
A Bayesian Approach to Systematic Error Correction}

\author{Jeffrey C. Smith\altaffilmark{*,1,2}, Martin C. Stumpe\altaffilmark{1,2},
    Jeffrey E. Van Cleve\altaffilmark{1,2}, Jon M. Jenkins\altaffilmark{1,2}, 
    Thomas S. Barclay\altaffilmark{1,3},
    Michael N. Fanelli\altaffilmark{1,3},
    Forrest R. Girouard\altaffilmark{1,4}, 
    Jeffery J. Kolodziejczak\altaffilmark{5}, 
    Sean D. McCauliff\altaffilmark{1,4},
    Robert L. Morris\altaffilmark{1,2}, 
    Joseph D. Twicken\altaffilmark{1,2}}

\altaffiltext{*}{email: jeffrey.smith@nasa.gov}
\altaffiltext{1}{NASA Ames Research Center, Moffett Field, CA 94035, USA}
\altaffiltext{2}{SETI Institute, 189 Bernardo Ave, Suite 100, Mountain View, CA 94043, USA }
\altaffiltext{3}{Bay Area Environmental Research Institute, 560 Third St West, Sonoma, CA 95476, USA}
\altaffiltext{4}{Orbital Sciences Corporation, 21839 Atlantic Blvd, Dulles, VA 20166, USA}
\altaffiltext{5}{Marshall Space Flight Center, Huntsville, AL 25812, USA}

%
%
%

\begin{abstract} 
With the unprecedented photometric precision of the Kepler Spacecraft, significant systematic and stochastic
errors on transit signal levels are observable in the Kepler photometric data~\citep{Jenkins:socOverview}.
These errors, which include discontinuities, outliers, systematic trends and other instrumental signatures,
obscure astrophysical signals.  The Presearch Data Conditioning (PDC) module of the Kepler data analysis
pipeline tries to remove these errors while preserving planet transits and other astrophysically interesting
signals. The completely new noise and stellar variability regime observed in Kepler data poses a significant
problem to standard cotrending methods such as SYSREM~\citep{tamuzMNRAS356} and TFA~\citep{kovacsMNRAS356}.
Variable stars are often of particular astrophysical interest so the preservation of their signals is of
significant importance to the astrophysical community.  We present a Bayesian Maximum A Posteriori
(MAP)~\citep{Kay:signalProcessing} approach where a subset of highly correlated and quiet stars is used to
generate a cotrending basis vector set which is in turn used to establish a range of ``reasonable'' robust fit
parameters. These robust fit parameters are then used to generate a \emph{Bayesian Prior} and a \emph{Bayesian
Posterior} Probability Distribution Function (PDF) which when maximized finds the best fit that simultaneously
removes systematic effects while reducing the signal distortion and noise injection which commonly afflicts
simple least-squares (LS) fitting. A numerical and empirical approach is taken where the Bayesian Prior PDFs
are generated from fits to the light curve distributions themselves.  
\end{abstract}


\section{An Overview of the Kepler Data Pipeline}\label{s:SOC}

\emph{Kepler's} primary science objective is to determine the frequency of Earth-size planets transiting their
Sun-like host stars in the habitable zone\footnote{The habitable zone is defined as the range of orbital
distances for which liquid water would pool on the surface of a terrestrial planet such as Earth, Mars, or
Venus without greenhouse gas adjustments to the atmosphere.}. This daunting task demands an instrument capable of measuring the light output from each of over
100,000 stars simultaneously with an unprecedented photometric precision of 20 parts per million (ppm) at
6.5-hour intervals. The large number of stars is required because the probability of the geometrical alignment
of planetary orbits that permit observation of transits is the ratio of the size of the star to the size of
the planetary orbit. For Earth-like planets in 1-Astronomical Unit (AU) orbits
about Sun-like stars, only
$\sim$0.5\% will exhibit transits. By observing such a large number of stars, \emph{Kepler} is guaranteed to
produce a robust result in the happy event that many Earth-size planets are detected in or near the
habitable zone. 

The Kepler Data Pipeline is divided into several components in order to allow for efficient management and
parallel processing of data. Raw pixel data downlinked from the \emph{Kepler} photometer are calibrated by the
Calibration module (CAL) to produce calibrated target and background pixels~\citep{CAL2010SPIE} and their
associated uncertainties~\citep{POU2010SPIE}. The calibrated pixels are then processed by the Photometric Analysis
module (PA) to fit and remove sky background and extract simple aperture photometry from the
background-corrected, calibrated target pixels\footnote{In simple aperture photometry, the brightness of a
star in a given frame is measured by summing up the pixel values containing the image of the
star.}~\citep{PA2010SPIE}. PA also measures the centroid locations of each star in each
frame. The final step to produce light curves is performed in the Pre-search Data Conditioning module (PDC),
where signatures in the light curves correlated with systematic error sources from the telescope and
spacecraft, such as pointing drift, focus changes, and thermal transients are removed.  Additionally, PDC
identifies and removes Sudden Pixel Sensitivity Dropouts (SPSDs) which result in abrupt drops in pixel flux
with short recovery periods up to a few hours, but usually not to the same
flux level as before. These step discontinuities are identified separately from those due to operational
activities, such as safe modes and pointing tweaks, and are mended using a sophisticated method described in a
companion paper \citep{Kolodziejczak:SPSD}. PDC also identifies residual isolated outliers and fills data
gaps (such as during intra-quarter downlinks) so that the data for each quarterly segment is contiguous when
presented to later pipeline modules. In a final step, PDC adjusts the light curves to account for excess flux in the
optimal apertures due to starfield crowding and the fraction of the target star flux in the aperture to make
apparent transit depths uniform from quarter to quarter as the stars move from detector to detector with each
roll maneuver.  Output data products include raw and calibrated pixels, raw and systematic error-corrected
flux time series, and centroids and associated uncertainties for each target star, which are archived to the
Data Management Center and made available to the public through the Multimission Archive at
STScI\footnote{http://stdatu.stsci.edu/kepler/}~\citep{FS2010SPIE}.  A companion paper describes the details
of overall PDC architecture~\citep{stumpe}.

Data is then passed to the Transiting Planet Search module (TPS)~\citep{TPS2010SPIE} where a wavelet-based
adaptive matched filter is applied to identify transit-like features with durations in the range of 1 to 16
hours. Light curves with transit-like features whose combined signal-to-noise ratio (SNR) exceeds 7.1$\sigma$
for a specified trial period and epoch are designated as Threshold Crossing Events (TCEs) and subjected to
further scrutiny by the Data Validation module (DV).  DV performs a suite of statistical tests to evaluate the
confidence in the transit detection, to reject false positives by background eclipsing binaries, and to
extract physical parameters of each system (along with associated uncertainties and covariance matrices) for
each planet candidate~\citep{DV2010SPIE,DVfitter2010SPIE}. After the planetary signatures are fitted, DV
removes them from the light curves and searches over the residual time series for additional transiting
planets. This process repeats until no further TCEs are identified. The DV results and diagnostics are then
furnished to the Science Team to facilitate disposition by the Follow-up Observing Program
(FOP)~\citep{gautierFOP2010}.

\section{A Bayesian Approach to Correcting Systematic Errors}\label{s:MAP}

\emph{Kepler} is opening up a new vista in astronomy and astrophysics and is operating in a new regime where
the instrumental signatures compete with the minuscule signatures of terrestrial planets transiting their host
stars. The dynamic range of the intrinsic stellar variability observed in the \emph{Kepler} light curves is
breathtaking:  RR Lyrae stars explosively oscillate with periods of approximately 0.5 days, doubling their
brightness over a few hours. Some flare stars double their brightness on much shorter time scales at
unpredictable intervals. At the same time, some stars exhibit quasi-coherent oscillations with amplitudes of
50 ppm that can be seen by eye in the raw flux time series \citep{jenkinsLC2010ApJ}. The richness of
\emph{Kepler's} data lies in the huge dynamic range for the variations in intensity by 4 orders of magnitude
and the range of time scales probed by the data, from a few minutes for SC data to weeks,
months, and ultimately, to years. Given that \emph{Kepler} was designed to be capable of resolving small
100-ppm changes in brightness over several hours, it is remarkably rewarding that it is revealing so much
more. The challenge is that an instrument so sensitive to the amount of light from a star striking a small
collection of pixels is also very sensitive to small changes in its environment.

The systematic errors observed in \emph{Kepler} light curves exhibit a range of different time scales, from a
few hours to several days to many days and weeks. Such phenomena include, for example, temperature variations
of the reaction wheel housing over the 3 day momentum managements cycles
and the resultant focus changes of $\sim$2.2 $\mu$m per $^\circ$C. Large thermal effects can be
observed in the science data for $\sim$5 days after recovering from intermittent safe modes, and for $\sim$3
days after attitude changes required to downlink the data each month which is due to different sides of the
spacecraft being heated during downlinks and subsequent thermal recoveries.  Another prominent systematic is
Differential Velocity Aberration (DVA) and the orbital period which results in gradual trends in the data
over each quarter. The principle objective of PDC is to remove these systematic effects by
\emph{cotrending}\footnote{ Detrending is the removal 
of low-frequency signal content regardless 
of origin (intrinsic or systematic). In contrast, \emph{cotrending} is the removal of 
signal content common to multiple targets, which can better
preserve intrinsic low-frequency signals while removing wide-band systematic signals.}.
The fact that most systematics such as these affect all the science data simultaneously, though
to differing degrees, provides significant leverage in dealing with these effects.

\subsection{The basic problem and the principle behind the solution} 

It is standard practice when removing systematic errors in stellar data to use robust least-squares (LS) on a
set of basis vectors as is used in methods such as SYSREM~\citep{tamuzMNRAS356} and
TFA~\citep{kovacsMNRAS356}.  A robust least-squares approach, as outlined below in section~\ref{ss:MAP},
can find a chance linear combination of the systematic error model components that reduces the bulk Root Mean
Square (RMS) at the expense of distorting the intrinsic stellar variations and introducing additional noise on
short timescales.  The fundamental problem with this approach is the fact that the implicit model fitted to
the data for each star is incomplete. Least-squares cotrending projects the data vector onto the selected
basis vectors and removes the components that are parallel to any linear combination of the basis vectors.
This process is guaranteed to reduce the bulk RMS residuals, but may do so at the cost of injecting additional
noise or distortion into the flux time series.  Indeed, this occurs frequently for stars with high intrinsic
variability, such as RR Lyrae stars, eclipsing binaries, and classical pulsators.  For example, if one of the
model terms is strongly related to focus variations and the long-term trend is for the width of the stellar
point spread function (PSF) to broaden over the observation interval, then the flux for all stars should
decrease over time. A least-squares fit, however, may invert the focus-related model term for a star whose
flux increases over the observation interval, thereby removing the signature of intrinsic stellar variability
from this light curve because there is a \emph{coincidental} correlation between the observed change in flux
and the observed change in focus. Given that the star would be expected to dim slightly over time, if
anything, due to the focus change, PDC should be correcting the star so that it brightens slightly more than
the original flux time series would indicate.  

The situation is analogous to opening a jigsaw puzzle box and finding only 30\% of the pieces present.
Least-squares
gamely tries to put the jigsaw pieces together in order to match the picture on the box cover by stretching,
rotating, and translating the pieces that were present in the box. The result is a set of pieces that roughly
overlap the picture on the box cover, but one where the details don't necessary match up well, even though
individual pieces may obviously fit. In order to improve the performance of robust LS, we need to provide the
fitter with constraints on the magnitudes and signs of the fit coefficients.  These constraints can be
obtained by using the ensemble behavior of the stars to develop an \emph{empirical} model of the underlying
physics. For example, the photometric change that can be induced by a pointing change of 0.1 arcsec must be
bounded, and this bound can be estimated by looking at how the collection of stars behaves for a pointing
change of this magnitude. 

\begin{figure}
\epsscale{1.0}
\plotone{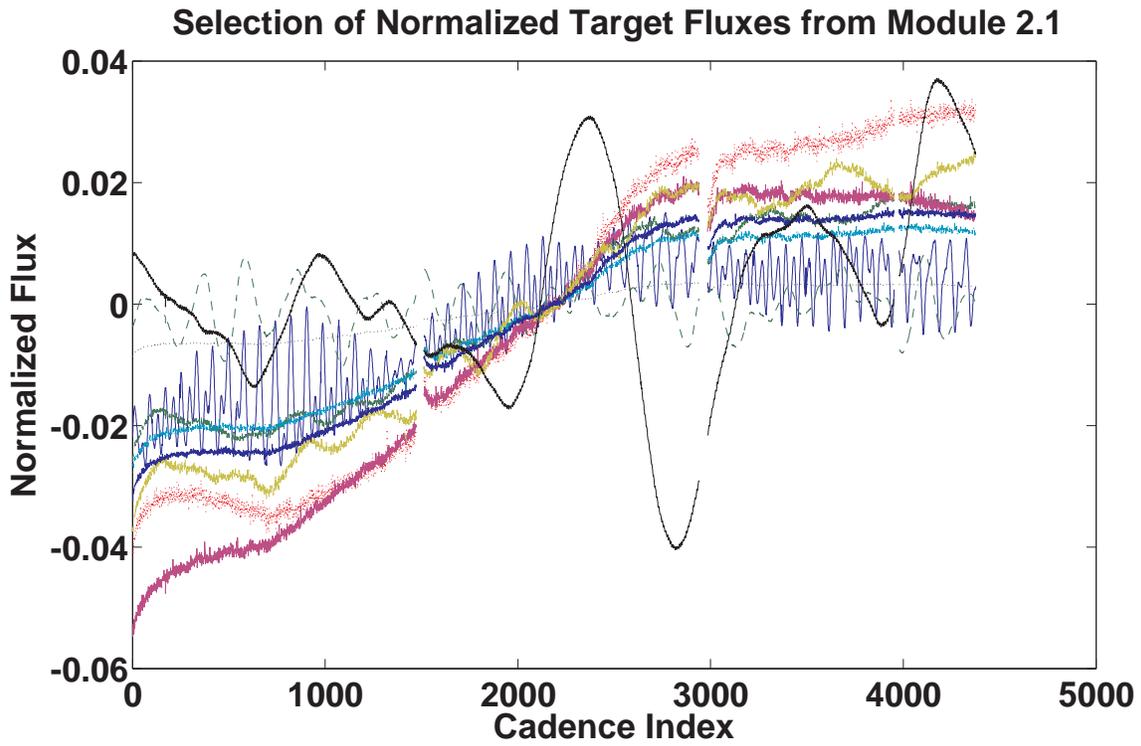}
\caption{Selection of typical light curves on channel 2.1. Notice the long term trend exhibited in all light
curves. However, for highly variable targets the trend is not entirely clear. This illustrates the need to
separate intrinsic stellar variability from systematic trends.}
\label{fig:typicalStarsOn2.1}
\end{figure}

\begin{figure}
\epsscale{1.0}
\plotone{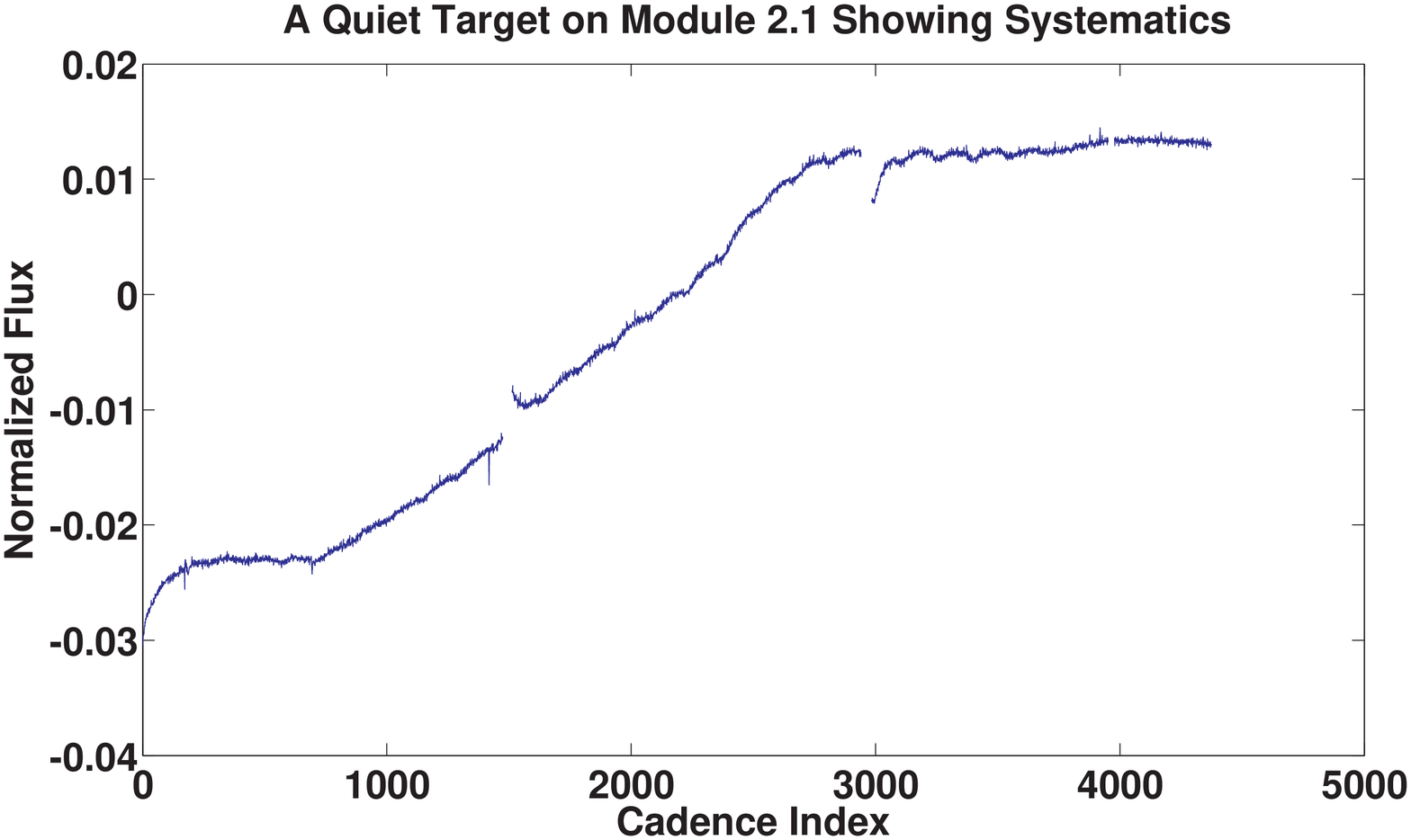}
\caption{A particularly quiet target on channel 2.1 showing almost purely systematic trends. The long term
trend is due to the seasonal changes to the shape of the telescope as the sun rotates around the barrel. Other
spacecraft systematics are also visible such as monthly Earth-point downlinks and heater cycling. Data gaps
and their thermal recoveries during the
monthly downlinks are evident at cadences 1500 and 2800.}
\label{fig:veryQuietStar}
\end{figure}

\begin{figure}
\epsscale{1.0}
\plotone{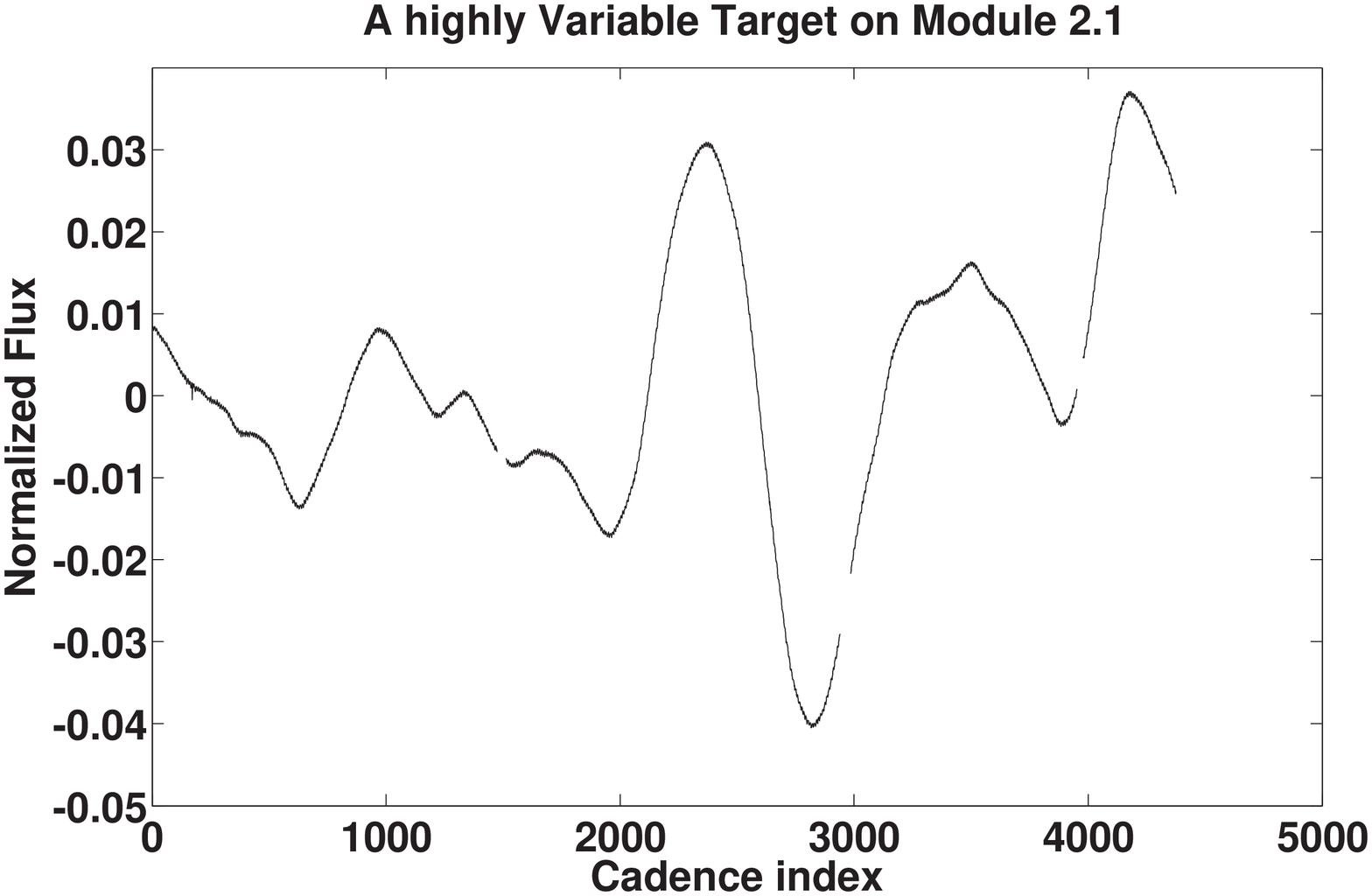}
\caption{A highly variable target where the variability completely obscures the systematic trend.}
\label{fig:veryNoisyStar}
\end{figure}

As an example of this analysis and to demonstrate systematic trends in the \emph{Kepler} 
data, take channel 2.1 which is the most thermally
sensitive CCD channel in \emph{Kepler's} focal plane. Nearly all stars on this channel exhibit obvious focus-
and pointing-related instrumental signatures in their pixel time series and flux time series. Figure
\ref{fig:typicalStarsOn2.1} shows several characteristic light curves for typical targets\footnote{The light
curves are referred to as ``targets'' and not stars since not all objects in the Kepler FOV are stellar.
Galactic studies are also performed with Kepler Data.} on channel 2.1 during the \emph{Kepler} 
Quarter 7 data season normalized by the median flux value. Note the long-term increase for all flux curves over the 90-day interval. 
This is due to seasonal changes in the shape of the telescope and therefore its focus as the Sun
rotates about the barrel of the telescope while the spacecraft orbits the Sun and maintains its attitude fixed
on the Field Of View (FOV). All light curves exhibit these long term trends but to differing degrees.
Also present are some short-term oscillations evident but mainly obscured in variable targets, which are due to focus changes
driven by a heater cycling on and off to condition the temperature of the box containing reaction wheels 3 and
4 on the spacecraft bus. This component was receiving more and more shade throughout this time interval and
the thermostat actuated more frequently over time. A target that is varying on levels and periods similar to
these systematic effects can obscure the systematics making identification difficult.  
Looking at a single quiet target shown in Figure~\ref{fig:veryQuietStar} (the same target shown in solid blue in
Figure~\ref{fig:typicalStarsOn2.1}) allows us to more clearly see the 
systematic trends 
which are exhibited in all the targets in Figure~\ref{fig:typicalStarsOn2.1} but obscured by variability. Each
Earth Point, one at the beginning of the quarter (cadence index 0) and after each monthly downlink 
(cadences ~1500 and ~2800) results in a heating of different sides of the telescope as the spacecrafts reorients
the antennae to downlink data. The Earth Points themselves are gaps in the data.
They result in periods of local heating and cooling distorting the telescope.
A characteristic recovery time is also evident. The other trends as described above are also clearly evident.
A final short data gap is also evident at cadence 3950, but this was not due to a reorientation of the
spacecraft so no thermal recovery is present.
As a counter example, Figure~\ref{fig:veryNoisyStar} shows the same highly variable target in solid black in
Figure~\ref{fig:typicalStarsOn2.1}. Notice how this highly
variable star almost completely obscures the long term trend. The targets shown in 
Figures~\ref{fig:veryQuietStar} and \ref{fig:veryNoisyStar} will hereby be referred to as the \emph{Quiet
Target} and the \emph{Variable Target} and used as canonical example targets in section~\ref{s:empBayesian}.

How can we separate intrinsic stellar variability from instrumental signatures? We do not expect intrinsic
stellar variability to be correlated from target to target, except for rare coincidences, and even then one would
not expect a high degree of correlation for all time scales. However, we \emph{do} expect instrumental
signatures to be highly correlated from target to target and can exploit this observation to provide constraints
on the co-trending that PDC performs. 

\begin{figure}
\epsscale{1.0}
\plotone{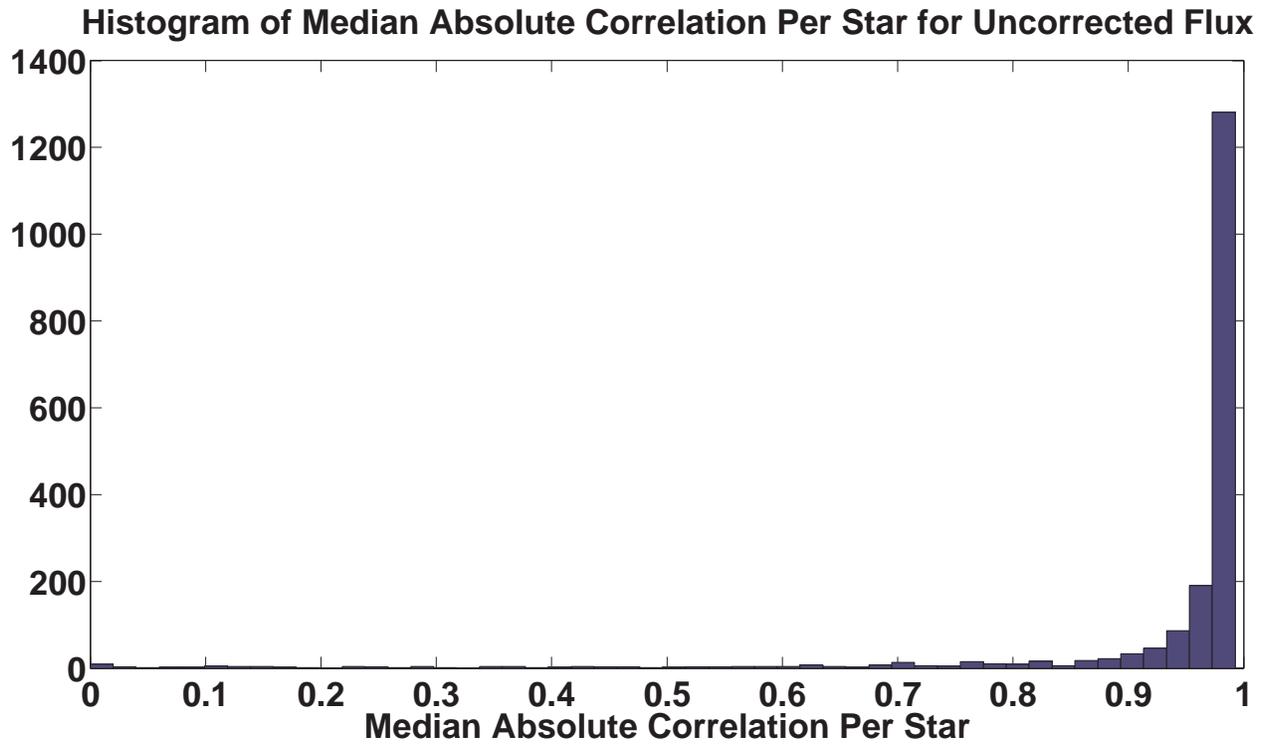}
\caption{Median Absolute Correlation for all targets on channel 2.1 Quarter 7.}
\label{fig:rawCorrelationHist}
\end{figure}

Figure \ref{fig:rawCorrelationHist} shows a histogram of the absolute value of the correlation coefficient for
1864 targets on channel 2.1. The targets' light curves are highly correlated as evidenced by the near complete pile-up near an
absolute correlation coefficient of 1. Examination of individual light curves indicates that these light
curves are contaminated to a large degree by instrumental signatures, as evidenced in Figures
\ref{fig:typicalStarsOn2.1} and \ref{fig:veryQuietStar}. But not all the targets are dominated by systematic errors. 
The trick
is to come up with a method that can distinguish between intrinsic stellar variability and chance correlations
with linear combinations of the diagnostic time series used to co-trend out systematic errors. 

\subsection{The MAP Approach, An Analytical Solution}\label{ss:MAP}

\newcommand{\Hmatrix}{\ensuremath{\mathbf{H}}}
\newcommand{\Cmatrix}{\ensuremath{\mathbf{C}}}
\newcommand{\Umatrix}{\ensuremath{\mathbf{U}}}
\newcommand{\Smatrix}{\ensuremath{\mathbf{S}}}

\newcommand{\thetavector}{\ensuremath{\boldsymbol{\theta}}}
\newcommand{\muvector}{\ensuremath{\boldsymbol{\mu}}}
\newcommand{\yvector}{\ensuremath{\mathbf{y}}}
\newcommand{\xvector}{\ensuremath{\mathbf{x}}}
\newcommand{\wvector}{\ensuremath{\mathbf{w}}}
\newcommand{\argmax}[1]{\underset{#1}{\mathrm{arg}\:\mathrm{max}}\:}
\newcommand{\pdf}[1]{p\left(#1\right)}
\newcommand{\thetaSub}[1]{\thetavector_{\mathbf{#1}}}
\newcommand{\hatThetaSub}[1]{\hat{\thetavector}_{\mathbf{#1}}}

A Bayesian approach called the Maximum A Posteriori (MAP) method allows us to provide PDC with
constraints on the fitted coefficients to help prevent over-fitting and distortion of intrinsic stellar
variability. In this exposition we follow the notation of \citep{Kay:signalProcessing}.

The PDC-MAP technique examines the behavior of the robust least-squares fit coefficients across an ensemble of
targets on each CCD readout channel in order to develop a description for the ``typical'' value for each model
term. This description is a probability density function (PDF) that can be used to constrain the coefficients
fitted in a second pass. To develop this approach, we build on a maximum likelihood approach. 

The maximum likelihood approach models each light curve, $\yvector$, as a linear combination of instrumental systematic
vectors, referred to as \emph{Cotrending Basis Vectors} or CBVs, arranged as the columns of a design matrix, \Hmatrix, plus zero-mean, Gaussian observation noise,
\wvector:

\begin{equation}
\displaystyle
\boldsymbol{\hat{y}} = \Hmatrix\thetavector + \wvector. 
\label{eq:dataModel}
\end{equation}
The Maximum Likelihood Estimator (MLE) seeks to find the solution, $\hat{\thetavector}_{\mathrm{MLE}}$, that
maximizes the likelihood function, $\pdf{\yvector; \thetavector}$, given by \begin{equation}
p\left(\yvector ; \thetavector\right) =  \frac{1}{ {\left(2\pi \right)}^\frac{N}{2} \left| \Cmatrix_{\wvector} \right|^{\frac{1}{2}}}
\exp \left[ -\frac{1}{2} \left(\yvector - \Hmatrix \thetavector \right)^T \Cmatrix_{\wvector}^{-1} \left(\yvector - \Hmatrix \thetavector \right)  \right],
\label{eq:MLE}
\end{equation}
where $\Cmatrix_{\wvector}$ is the covariance of $\wvector$ and $N$ is the number of data points.  Taking the gradient of the log of Equation
\ref{eq:MLE}, setting it equal to zero, and solving for \thetavector\ yields the familiar least-squares
solution,
\begin{equation}
\hatThetaSub{MLE} = \left( \Hmatrix^T \Cmatrix_{\wvector}^{-1} \Hmatrix\right)^{-1}
\Hmatrix^T \Cmatrix_{\wvector}^{-1}\yvector.
\label{eq:MLEsolution}
\end{equation}
This solution assumes the model $\Hmatrix$ is a complete model to the data. We will show that the Bayesian model
accounts for an incomplete model which is the common case when removing systematics from stellar signals.



Adopting the Bayesian approach allows us to incorporate side information, such as knowledge of prior
constraints on the model, in a natural way. Bayesianists view the underlying model as being drawn from a
distribution and the data as being one realization of this process. In this case we wish to find the Maximum A
Posteriori (MAP) estimator of the model coefficients given the observations (data):
\begin{equation}
\hatThetaSub{MAP} = \argmax{\thetavector} p(\thetavector | \yvector) =\argmax{\thetavector} \pdf{\yvector | \thetavector} \pdf{\thetavector},
\label{eq:MAPdefinition}
\end{equation}
where we've applied Bayes' rule~\citep{dagonstini} to simplify the expression. In this equation,  $p\left(\thetavector \right)$
is the \emph{prior PDF} of the model coefficients. The mathematical form for $p(\yvector|\thetavector)$ is the same
as for the non-Bayesian likelihood function $\pdf{\yvector ; \thetavector}$ in Equation \ref{eq:MLE}. 

For illustration purposes, if we adopt a Gaussian form for the coefficient distribution, $\thetavector$,
then $p(\thetavector)$ takes a closed form solution, 
\begin{equation}
p(\thetavector) =  \frac{1}{ \left(2\pi \right)^\frac{M}{2} \left| \Cmatrix_{\thetavector} \right|^\frac{1}{2} }
\exp \left[ -\frac{1}{2} \left( \thetavector - \muvector_{\thetavector} \right)^T \Cmatrix_{\thetavector}^{-1} 
\left(\thetavector - \muvector_{\thetavector} \right)  \right],
\label{eq:gaussianPrior}
\end{equation}
where $\Cmatrix_{\thetavector}$ and $\muvector_{\thetavector}$ are the covariance and mean of \thetavector, respectively,
and we assume that the coefficients are uncorrelated (which will hold true for orthogonal basis functions), 
we can then maximize Equation \ref{eq:MAPdefinition}, using Equation~\ref{eq:gaussianPrior}, 
by maximizing its log likelihood,
\begin{multline}
\ln [ p(\yvector|\thetavector) \: p(\thetavector ) ] = 
\ln [ p(\yvector|\thetavector) ] + \ln [ p(\thetavector ) ] \\
= -\frac{N}{2} \ln \left( 2 \pi \right) - \frac{1}{2} \ln \left| \Cmatrix_{\wvector} \right| 
 -\frac{1}{2} \left(\yvector - \Hmatrix \thetavector \right)^T \Cmatrix_{\wvector}^{-1} \left(\yvector - \Hmatrix \thetavector \right) \\
 -\frac{M}{2} \ln \left( 2 \pi \right) - \frac{1}{2} \ln \left| \Cmatrix_{\thetavector} \right| 
 -\frac{1}{2} \left( \thetavector - \muvector_{\thetavector} \right)^T \Cmatrix_{\thetavector}^{-1} 
\left(\thetavector - \muvector_{\thetavector} \right).
\label{eq:logLikelihoodMAP}
\end{multline}
Taking the gradient of equation~\ref{eq:logLikelihoodMAP} with respect to \thetavector, 
setting it to zero, and solving for \thetavector\ yields
\begin{equation}
\displaystyle
\hatThetaSub{MAP} = 
\left( \Hmatrix^{T} \Cmatrix_{\wvector}^{-1} \Hmatrix + \Cmatrix_{\thetavector}^{-1} \right )^{-1}
\left( \Hmatrix^T \Cmatrix_{\wvector}^{-1} \yvector + \Cmatrix_{\thetavector}^{-1} \muvector_{\thetavector} \right).
\label{eq:fullMAPsolution}
\end{equation}
If the observation noise, $\wvector$, is zero-mean, white Gaussian noise with variance $\sigma^2$, 
then Equation \ref{eq:fullMAPsolution} can be rewritten as
\begin{equation}
\hatThetaSub{MAP} 
= \left( \Hmatrix^{T} \Hmatrix +  \sigma^{2}\Cmatrix_{\thetavector}^{-1} \right )^{-1}
\left( \Hmatrix^T \yvector + \sigma^{2} \Cmatrix_{\thetavector}^{-1} \muvector_{\thetavector} \right).
\label{eq:simplifiedMAPsolution}
\end{equation}

The key to this Bayesian technique is to determine when to preference, or weight, the prior PDF over the
conditional PDF.  If the variance in the data is large compared to the ``spread'' allowed by the prior PDF for
the model, then the MAP estimator gives more weight to the prior so that $\hatThetaSub{MAP}\rightarrow
\muvector_{\thetavector}$ as $\sigma^2 \rightarrow \infty$. This case would correspond, for example, to
targets with large stellar variability such as with the target given in Figure~\ref{fig:veryNoisyStar}. In
this case, the MAP weighting constrains the fitter from distorting the light curve and introducing noise on a
short time scale.  Conversely, if the variance in the data is small compared to the degree to which the prior
PDF confines the model, the MAP estimator ``trusts'' the data over the prior knowledge and
$\hatThetaSub{MAP}\rightarrow \hatThetaSub{MLE}$ as $\sigma^2 \rightarrow 0$.  This case would correspond to
targets with small stellar variability such as with the target in Figure~\ref{fig:veryQuietStar} where there
is little risk of over-fitting and distortion of the light curves and it is a ``safe'' bet to use the
conditional, least-squares fit.

\section{The Empirical Bayesian MAP Approach and Implementation}\label{s:empBayesian}

The above analytical solution to the Bayesian posterior PDF restricts the prior PDF to a Gaussian form.  There
is no a-priori reason to make this assumption and in general, since we are developing an \emph{empirical}
prior PDF, the least number of analytical constraints on the form, the more complete will be the empirical
model.

If a Gaussian form to the prior is no longer assumed then the prior formalism in
equation~\ref{eq:gaussianPrior} can no longer be used. We can, however, still take the log form of
equation~\ref{eq:MAPdefinition} to obtain
\begin{equation}
\hatThetaSub{MAP} = \argmax{\thetavector} p(\thetavector | \yvector) =\argmax{\thetavector} 
    \left( \log\left(\pdf{\yvector | \thetavector}\right) +  \log\left(\pdf{\thetavector}\right) \right).
\label{eq:MAPlogForm}
\end{equation}
Using the Maximum Likelihood Estimator in equation~\ref{eq:MLE} for $p(\yvector|\thetavector)$, 
removing the constant terms, inserting a weighting parameter and using normalized light curves, $\hat{y}$, we obtain
\begin{equation}
\hatThetaSub{MAP} = \argmax{\thetavector} 
    \left[ -\frac{1}{2 \sigma^{2}} \left( \hat{y} - \Hmatrix \thetavector \right)^{\mathrm{T}} \left( \hat{y} - \Hmatrix
    \thetavector \right) + \mathbf{W_{pr}} \log p(\thetavector)\right],
\label{eq:mapEmpiricalSoln}
\end{equation}
where we assume the observation noise, \wvector, is zero-mean, white Gaussian noise and has variance
$\sigma^2$. Since $p(\thetavector)$ is no longer in closed form, the ``spread'' in the prior PDF (i.e. the covariance
of $\thetavector$, $\Cmatrix_{\thetavector}$ in equation~\ref{eq:simplifiedMAPsolution}) can no longer be
expressed succinctly. In its stead, a \emph{generalized weighting parameter}, $\mathbf{W_{pr}}$, is used to characterize the 
``spread'' in the prior PDF.
Equation~\ref{eq:mapEmpiricalSoln} must now be evaluated numerically.

\begin{figure}
\epsscale{0.6}
\plotone{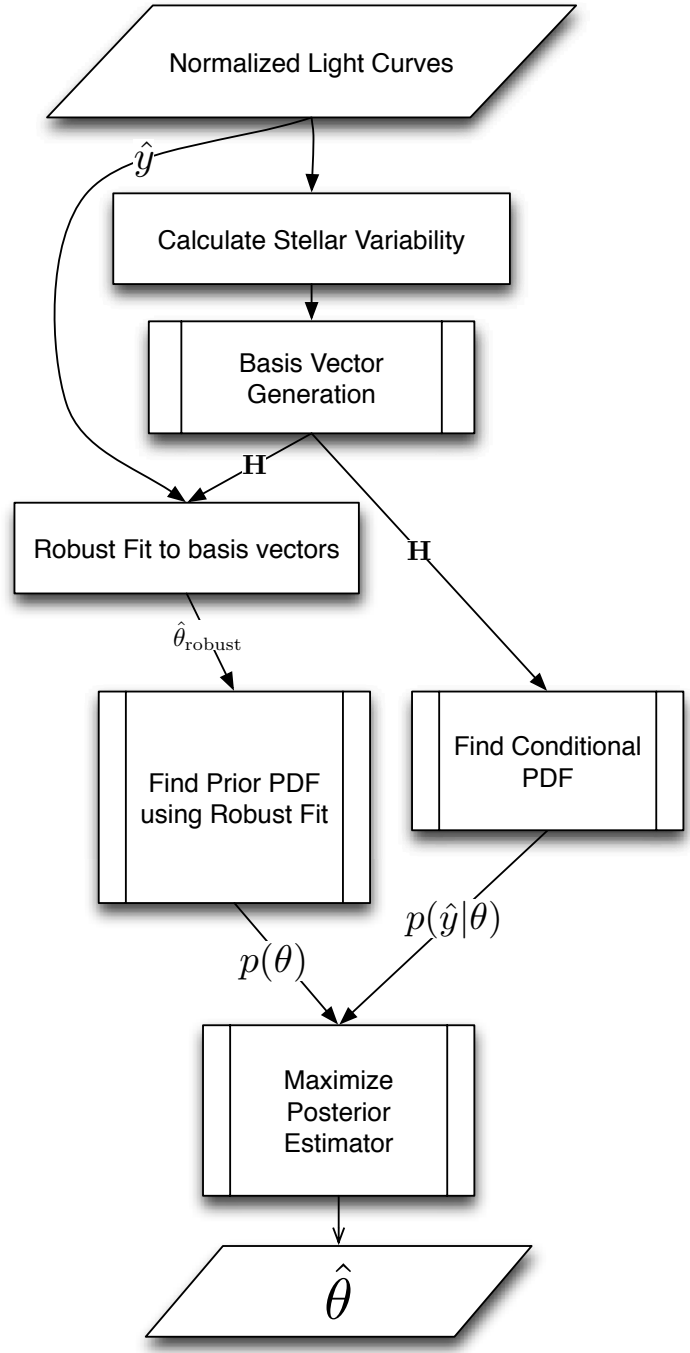}
\caption{Flowchart of the PDC-MAP cotrending algorithm.}
\label{fig:map_flowchart}
\end{figure}

The overall flow of the algorithm is shown in Figure~\ref{fig:map_flowchart}.  We start by normalizing the flux
light curves and calculating a relative stellar variability. We then find
basis vectors using SVD based on a reduced set of flux light curves where cuts are made on target-to-target
correlation and stellar variability. A robust least-squares fit is then performed on each target using the
basis vectors just found. This ensemble of \emph{robust fit coefficients} is used to generate the prior PDF. 
The conditional PDF is
also found based on the same basis vectors. Once both prior and conditional PDFs are found they are
combined to generate the posterior PDF where a weighting parameter, based on the stellar variability and the
``goodness'' of the prior fit, is used to weigh the prior relative to the conditional PDF. 
Details are elucidated in the following subsections.

\subsection{Finding the Cotrending Basis Vectors}
\label{s:findingBasisVectors}

The Cotrending Basis Vectors are obtained using Singular Value Decomposition.  In order to have equal
representation for all light curves independent of their absolute magnitude, we first normalize the targets
by their median flux values ($\frac{\Delta \mathrm{flux}}{\mathrm{median}(\mathrm{flux})}$). We then select the 50\% most
highly correlated targets based on the median absolute Pearson correlation. This cut generates a set that
exhibits the strongest trends in the data. It mostly removes targets with large variability but not
completely. A variable star exhibiting a strong trend can still remain in the reduced list. We therefore first
make a cut on the estimated variability of each target.

\begin{figure}
\epsscale{1.0}
\plotone{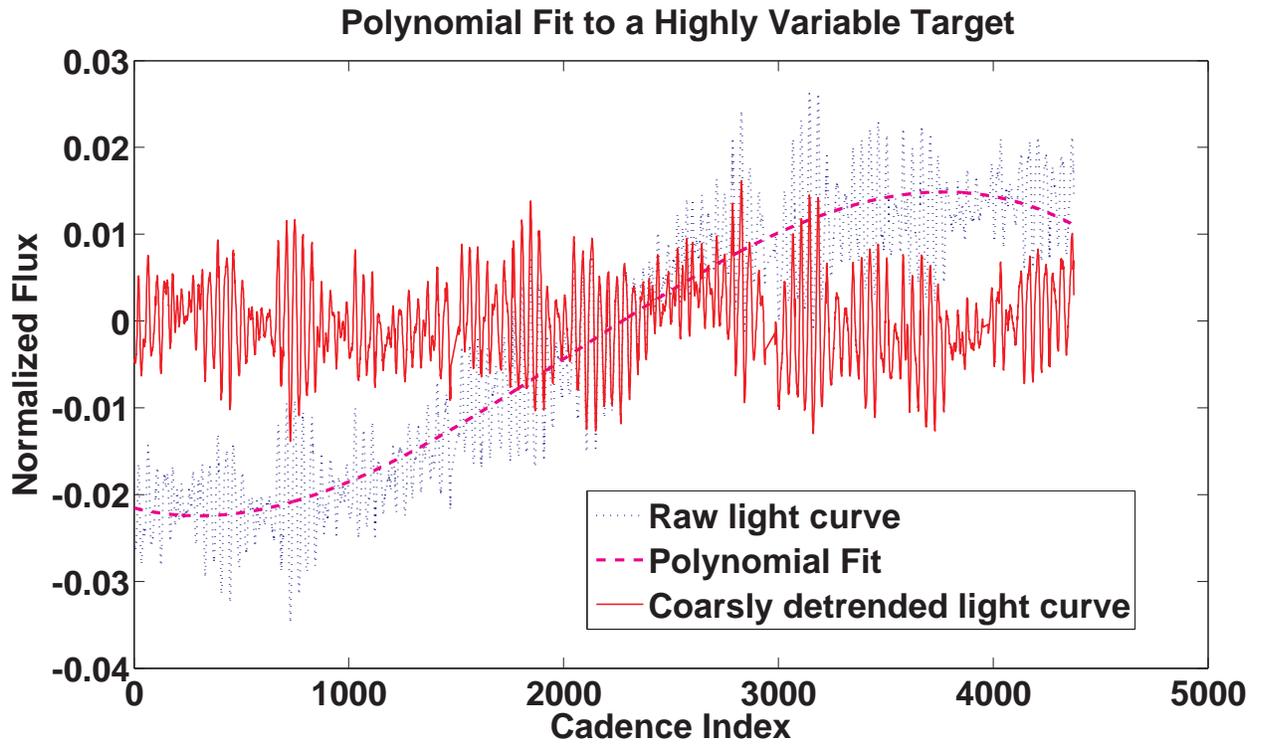}
\caption{By removing a 3rd order polynomial fit to the raw light curve an estimate of the intrinsic
variability of the target can be calculated.}
\label{fig:variabilityFit}
\end{figure}

An estimate of the intrinsic stellar variability of each target must be found. Herein lies the fundamental
chicken-and-egg problem of the cotrending method. We need to know the stellar variability of each target in
order to know how much to rely on the prior. But if we already knew the stellar variability then we would have
no need for the prior -- the cotrending solution would simply be the intrinsic stellar variability subtracted
from the light curve, minus a Gaussian noise estimate. This issue is not specific to this particular
cotrending method either. Whenever a system is characterized with an incomplete model there exists the problem
of identifying the components in the system not represented in the model.  We fortunately do not need to
absolutely know the variability, we only need an estimated metric in order to weigh the prior. This estimate
can be obtained by comparing a third-order polynomial to the light curve. The polynomial will remove any long
term trends leaving behind a \emph{roughly} detrended curve.  The standard deviation of this polynomial
removed light curve results in a rough calculation of the variability of the target.  Removing a low-order
polynomial is essentially a high-pass filter, we are therefore assuming any long term trends are systematic
and short term trends are stellar. There are numerous counter-examples of short term trends that are actually
systematic -- reaction wheel zero crossingsn is a good example. However, short term systematic trends tend to
be small in magnitude whereas long term systematics tend to result in large diversions in the flux amplitude.
Likewise, there are examples of intrinsic long-term trends but they are generally smaller than the systematic
trends. Since we are only concerned with the \emph{relative} amplitude of stellar versus systematic
variability, we are using the low pass filter to distinguish two characteristic realms of influence: long term
trends dominated by systematics and short term trends dominated by intrinsic stellar variation.  An example is
shown in Figure~\ref{fig:variabilityFit}.  Here, a highly variable target is compounded with a long-term DVA
and thermal trend. For periods less than $400$ cadences the variance in the flux is dominated by stellar
features. The long term variance, and the general trend to higher flux values is due to systematics. The
variance of the residual after removing the polynomial fit, labeled as ``Coarsely detrended light curve'' in
the figure, gives a \emph{rough} estimate of the stellar variability of this target. Note that there are still
systematic features in the detrended light curve. They are however small in magnitude compared to the stellar
variability\footnote{This does not lessen the ability of PDC-MAP to remove short term systematics. Such short
term systematics are still present in the basis vectors and so when the PDF fit is performed the short term
trends are removed.}.

\begin{figure}
\epsscale{1.0}
\plotone{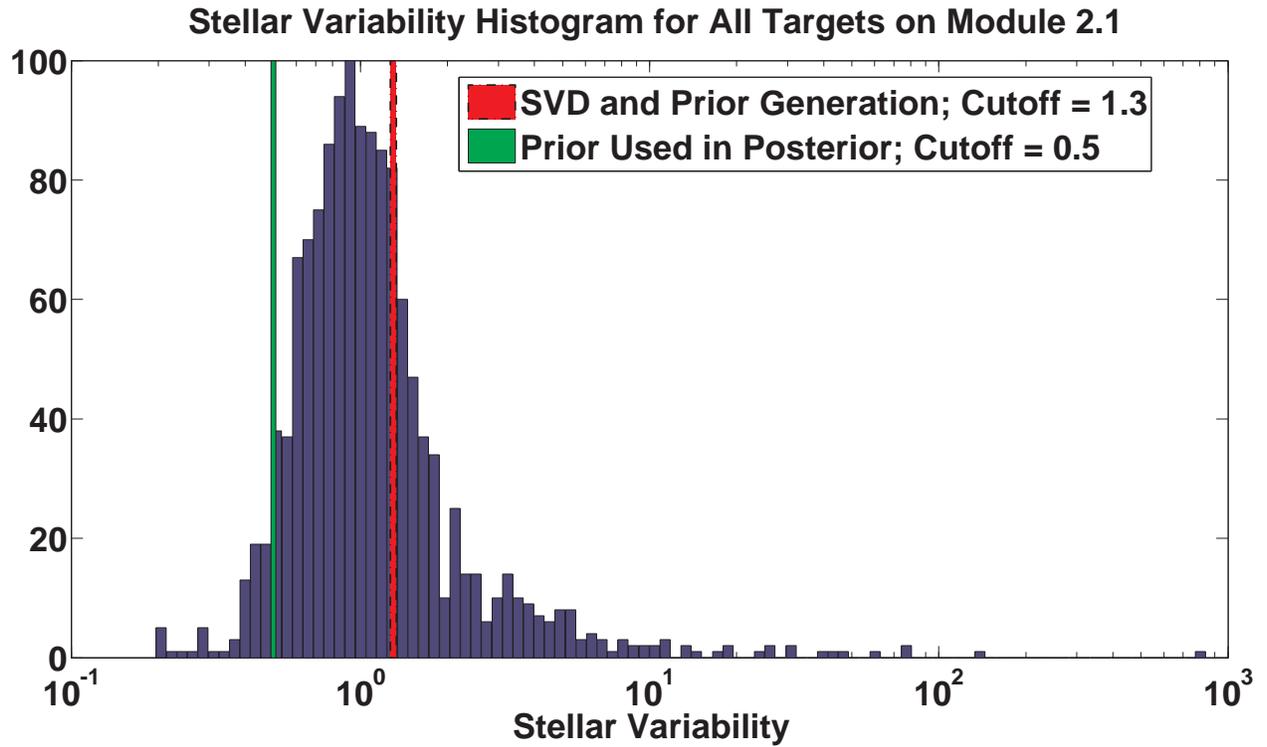}
\caption{Histogram of estimated variability for all targets on channel 2.1. This distribution is typical for all Kepler
channels. The quiet targets below the ``SVD and Prior Generation'' threshold are used to generate the cotrending basis vectors.}
\label{fig:variabilityScatter}
\end{figure}

The variability, $V$, is measured using
\begin{equation}
V =\frac{\sigma_{\hat{y}}} {\Delta y \tilde{V}},
\label{eq:stellarVariability}
\end{equation}
where $\sigma_{\hat{y}}$ is the standard deviation of the third-order polynomial detrended light curve,
$\Delta y$ is the uncertainty of the flux data as determined by the PA pipeline component~\citep{POU2010SPIE} 
and $\tilde{V}$ is the median variability over all light
curves in the sample. The normalization by the uncertainty is to ensure the noise in the data is not
included in the stellar variability. The normalization by the median variability is so that a variability of 1
is considered typical thereby simplifying the analysis parameterization.
Figure~\ref{fig:variabilityScatter} shows a histogram of the measured variability for all targets on channel 2.1. The median
of all of these values is evidently 1 and the distribution is typical for all channels where most are close to
typical variability but with a long tail to high variability (note the log scale for the x-axis). 
There are two cutoff thresholds plotted as well. The upper (in dashed red) is
the threshold to determine if a target is ``highly variable.''  The lower (in solid green) is to determine if a
target is ``very quiet.'' The very quiet targets have such a low amount of variability that using the prior PDF when
generating the fit has been found to be problematic.
Any targets above the high variability threshold are removed from the reduced list. 
The remaining targets are sorted
with respect to median absolute correlation and the 50\% most highly correlated are used for SVD. 

Due to all targets being normalized by their median, most targets pass through zero
amplitude at the midpoint as can be seen as a ``node'' in the light curves at cadence 2200 in
Figure~\ref{fig:typicalStarsOn2.1}. If SVD was performed on this set, as is, then all the strong cotrending
basis vectors would have zero amplitude at the midpoint. The basis vectors would therefore be unable to remove
systematics in the minority of targets that do not pass through zero at the midpoint. The light curves are
therefore \emph{dithered} slightly by a zero-mean Gaussian dithering magnitude in order to slightly ``spread'' the
light curves about the zero flux value. Since the dithering is zero mean this has no effect on the resultant
basis vectors other than to remove the artificial zero crossing node at the midpoint. Note that the dithering
is only used to generate the basis vectors. The cotrending is performed on the non-dithered, but still
median-normalized, light curves.  

\begin{figure}
\epsscale{1.0}
\plotone{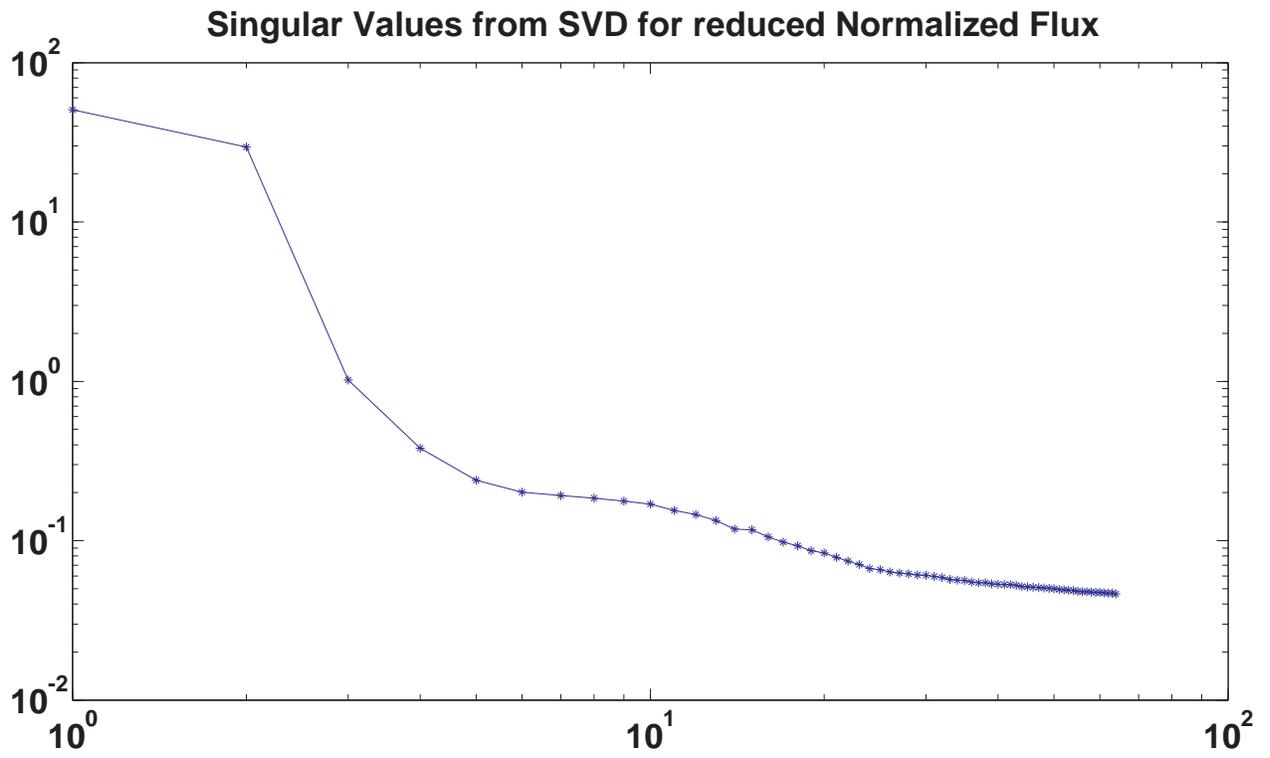}
\caption{The singular values from SVD on the reduced set of ``cleaned'' targets that are highly correlated and
quiet.}
\label{fig:singularValues}
\end{figure}

Figure~\ref{fig:singularValues} shows the singular values from the singular
value decomposition. This figure is characteristic of all channels; 2 or 3 strong singular values,
then a slowly tapering region for about another dozen values until finally asymptotically approaching zero (as
is expected with SVD).  The first several left singular vectors are selected (typically the first 8) to become
the \emph{Cotrending Basis Vectors}. These first singular vectors exhibit the principle trends in the
data due to DVA, pointing errors, impulses due to Argabrightenings~\citep{witteborn}, focus errors and reaction
wheel zero crossings among other trends. The number of basis vectors used is generally 8, however a
signal-to-noise ratio test is performed where the SNR is determined by
\begin{equation}
\mathrm{SNR_{db}} = 10 \log_{10} \left( \frac{A_{\mathrm{signal}}^{2}}{A_{\mathrm{noise}}^{2}}\right).
\end{equation}
$A_{\mathrm{signal}}$ and $A_{\mathrm{noise}}$ being the Root Mean Square of the light curve and noise floor
respectively. The noise floor is approximated by the first differences between adjacent flux values.
Any of the 8 basis vectors with a SNR below a threshold of 5 decibels are
removed but only a small number of basis vectors over the entire field of view are removed by the SNR test. 
Most have
high SNR. There are other sophisticated methods to find the dimensionality of an eigensystem such as Bayesian
Model Selection \citep{minka}. These are not used because they tend to pick too high a dimensionality in this
particular situation. We
wish to find only the singular vectors with systematics, the lesser singular vectors do contain light curve
signal information, but not necessarily systematics and we have found including them in the MAP fit adds no
value yet slows down the algorithm.

For a minority of basis vectors, a few target light curves can dominate the signal. The normalization
process attempts to ``equalize'' the strength of all targets, but a small number of light curves can be
over-represented in the singular vectors from SVD. To eliminate this we calculate an entropy metric for each
basis vector using the following entropy calculation
\begin{equation}
h(p_{i}) = - \int p(x) \log p(x) dx,
\end{equation}
where p(x) is a probability distribution function created from the right singular vectors from SVD (referred
to as the \emph{V-Matrix}),
\begin{equation}
p_{i}(x) = \{ \mathbf{V}_{ki} \}.
\end{equation}
The V-Matrix contains the contribution of the signal in the basis vector from each target light curve.  We must first
normalize the entropy calculation to a Gaussian distribution, which has the highest entropy of any continuous distribution
with the same 2nd moment. The entropy of a Gaussian is
\begin{equation}
H_{0}(\sigma) = \frac{1 + \log (2 \pi)}{2} + \log (\sigma),
\end{equation}
$\sigma$ being the 2nd central moment of $V_{ki}$ for fixed $i$. The resultant relative entropy is therefore
\begin{equation}
h'(p_{i}) = h(p_{i}) - H_{0}(\sigma).
\end{equation}
If one (or a few) targets are domineering then they will have much larger values in the V-Matrix then all the
other targets. A negative value of the entropy calculation will identify this condition.  Bad entropy is
somewhat arbitrary but we have found that a value below $-0.7$ is poor. For any basis vectors
with identified poor entropy, the V-matrix column for that basis vector is examined for stand-out targets.
The offending targets are removed and SVD is re-computed on the remaining targets. The process is iterated
until the entropy of all basis vectors is below $-0.7$. Typically, no more than a couple iterations is
necessary and fewer than 20 targets are removed (out of ~2500 total targets).

\begin{figure}
\epsscale{1.0}
\plotone{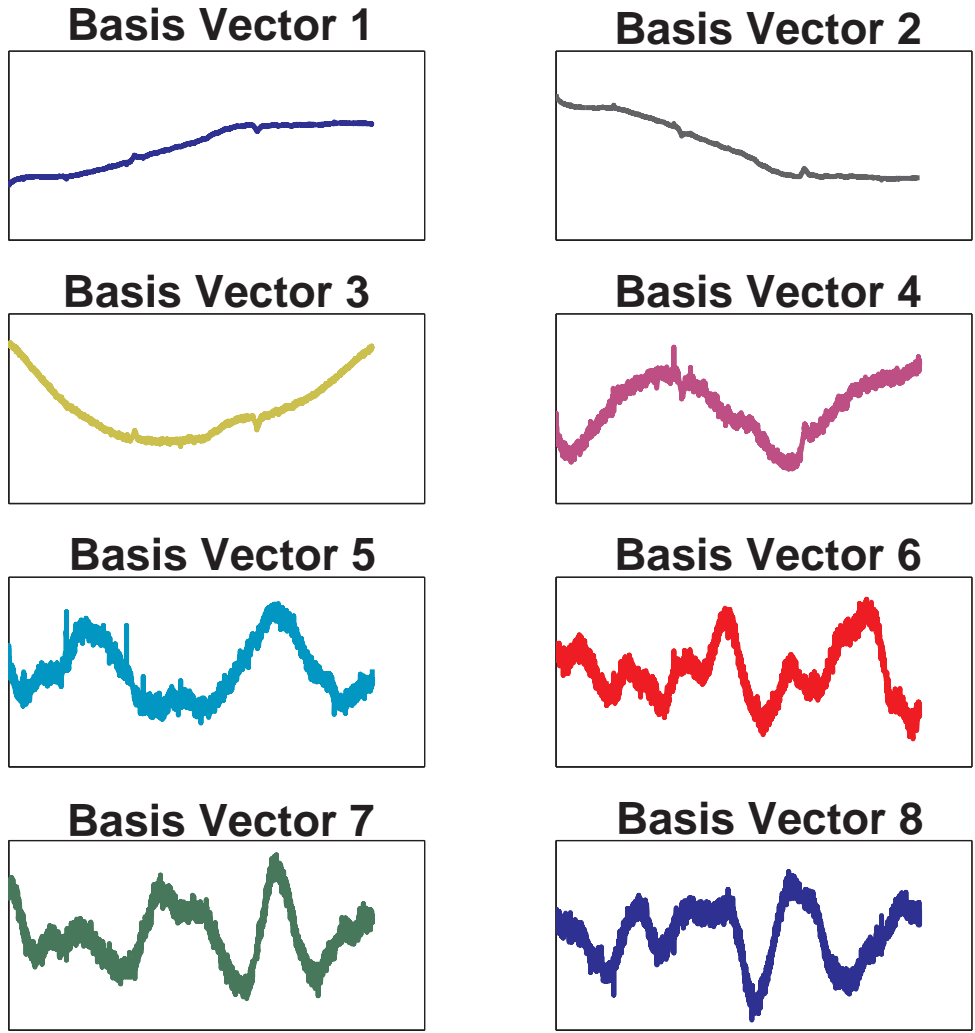}
\caption{First 8 Cotrending Basis Vectors for channel 2.1. The gaps in the data have been linearly filled so
these curves are continuous.}
\label{fig:basisVectors}
\end{figure}

\begin{figure}
\epsscale{1.0}
\plotone{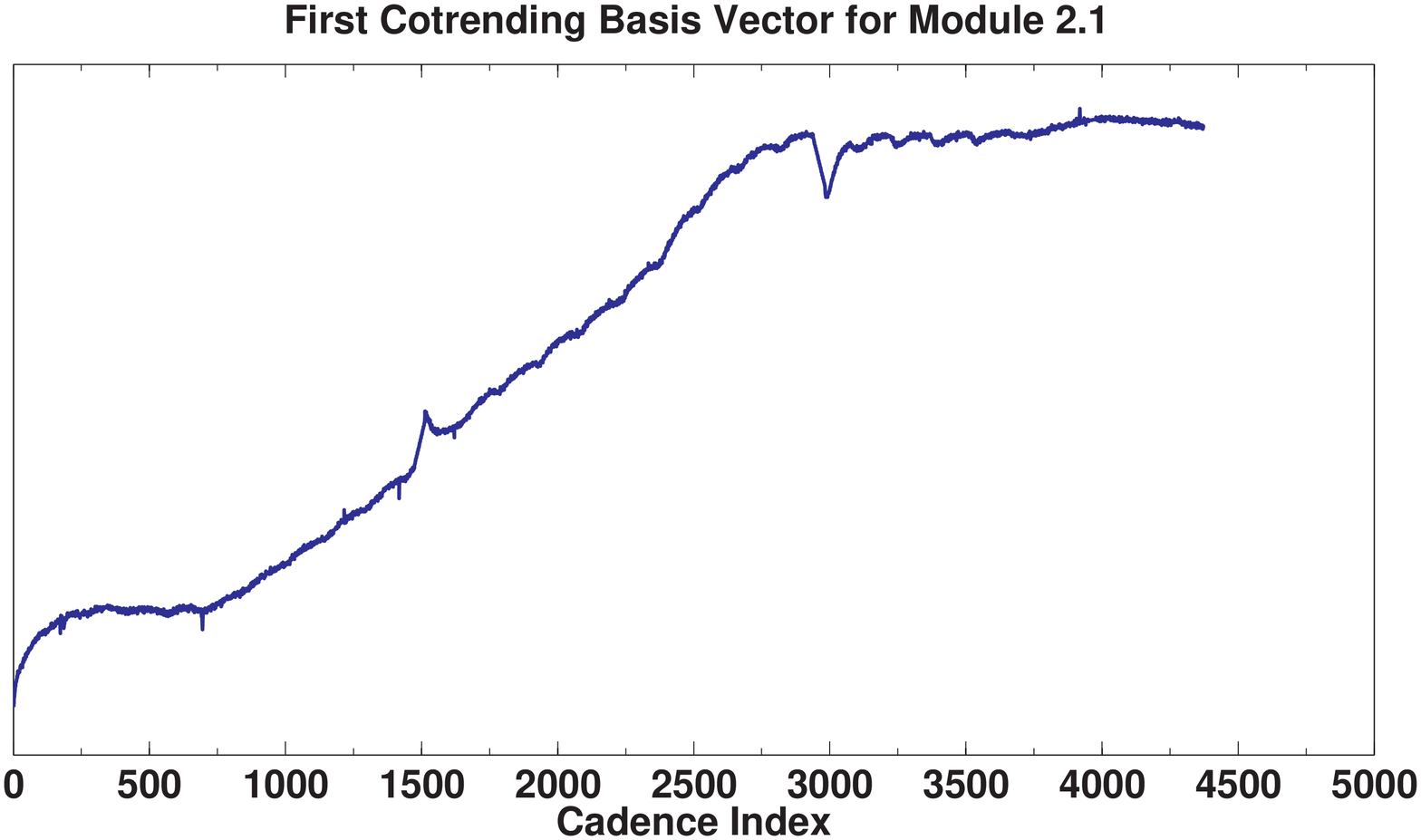}
\caption{First Cotrending Basis Vector for channel 2.1. The amplitude of the basis vector is arbitrary.}
\label{fig:firstBasisVector}
\end{figure}

Figure~\ref{fig:basisVectors} shows the first eight cotrending basis vectors generated for channel 2.1 and
Figure~\ref{fig:firstBasisVector} shows just the first basis vector.
Trends can be found in all the vectors
but it is useful to concentrate on the first, and strongest. Here the most characteristic trends and
systematics in the data can be found. The general trend to higher flux is due to the seasonal change
and solar orientation. The short recovery periods at cadence indices 0, 1500 and 2800 are due to monthly
downlinks. The short spikes at 700 and 1450 are due to artifacts from correcting cosmic rays near reaction
wheel zero crossing periods\footnote{These artifacts have been resolved in a recent version of the PA pipeline
component but Argabrightenings still persist.}. The periodic oscillation is due to heater
cycling. Notice how the Basis Vector in Figure~\ref{fig:firstBasisVector} closely follows the Flux Light Curve in
Figure~\ref{fig:veryQuietStar}. This signifies that virtually all the features in this Flux Light Curve are
due to systematic effects and not intrinsic stellar variability. In theory, any features in the light curves
in Figure~\ref{fig:typicalStarsOn2.1} that are not represented in the basis vectors in
Figure~\ref{fig:basisVectors} are intrinsic to the target. However, a simple least-squares projection of the
light curves on the basis vectors will not produce desirable results for all targets as will be shown below.

Once the cotrending basis vectors are found a robust LS fit is performed on each target. This creates the
empirical data used to generate the prior PDF.

\subsection{Numerically Generating $p(\thetavector)$}
\label{sec:findPrior}

The prior PDF is based on the distribution of robust fit coefficients of the basis vectors for all light
curves using the method described in~\citep{ksdensityReference}. This method computes a probability density
estimate of the sample data based on a normal kernel function using a window that is a function of the number
of points in the data sample. The form of the prior PDF will depend on the 
parameterization of the robust fit
coefficients. We must thus decide how to parameterize the coefficients to best extract the correlations.
Some systematic effects are caused by focal plane irregularities and instrumental vibrations
which are stronger near the edges of the CCD frame (nearer to the spacecraft housing). There are also other 
issues that are dependent on the physical position of each pixel on the CCD. Therefore, the
targets' locations in the sky as characterized by right ascension and declination are reasonable parameters to
characterize target location with respect to the sources of systematic effects. The targets' influence by
systematic effects is also directly related to the stellar magnitude since different magnitude targets result in
different saturation levels of the CCD pixels. For example, the readout electronics for the CCDs are sensitive
to temperature drift but the sensitivity is non-linear with respect to CCD flux levels. So brighter targets are
affected by instrument temperature differently than dim targets.  We therefore parameterize the prior PDF with
three independent variables: \begin{inparaenum} \item Stellar Magnitude ($K_{p}$) \item Right Ascension (RA)
and \item Declination (Dec) \end{inparaenum}.  

\begin{figure}
\epsscale{1.0}
\plotone{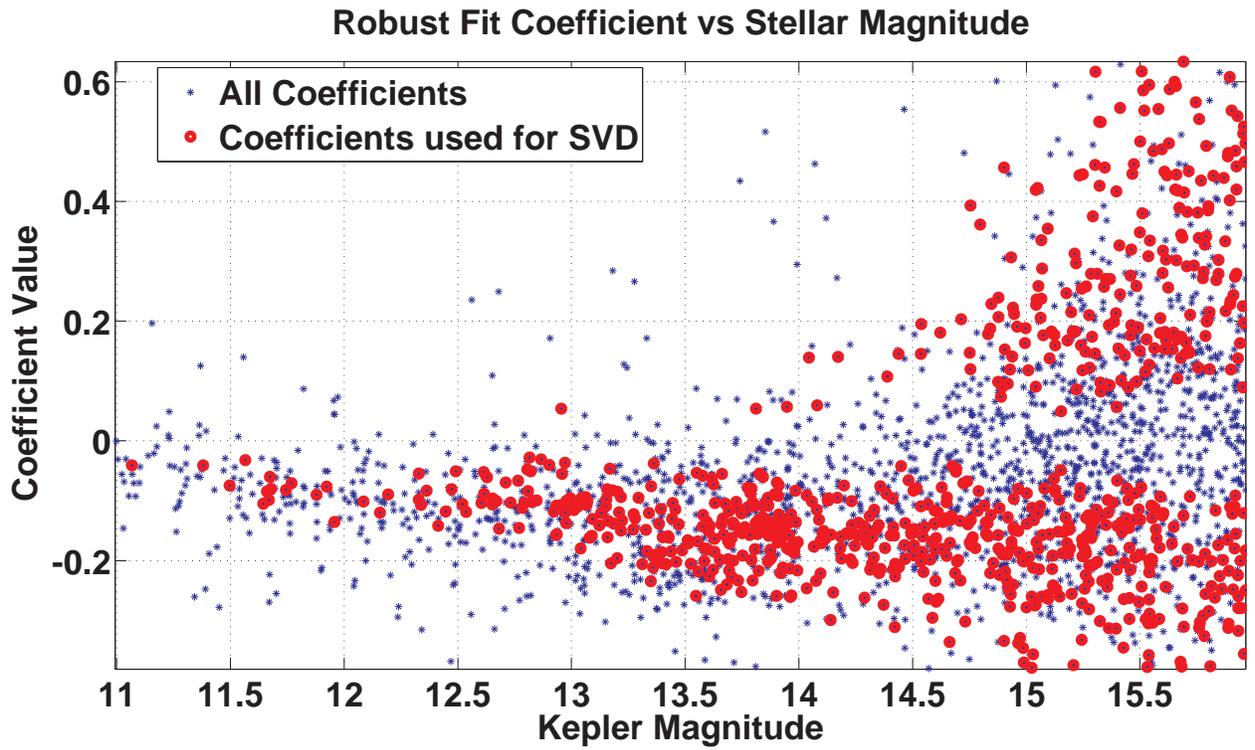}
\caption{Robust fit coefficients for Basis Vector 1 for all targets (Blue stars) and only those targets used for SVD
(Red circles) plotted against Kepler Magnitude. By taking cuts on stellar variability, target-to-target
correlation and entropy results in a bimodal distribution that would not be evident without the cuts.}
\label{fig:coeff2VsK_p}
\end{figure}

\begin{figure}
\epsscale{1.0}
\plotone{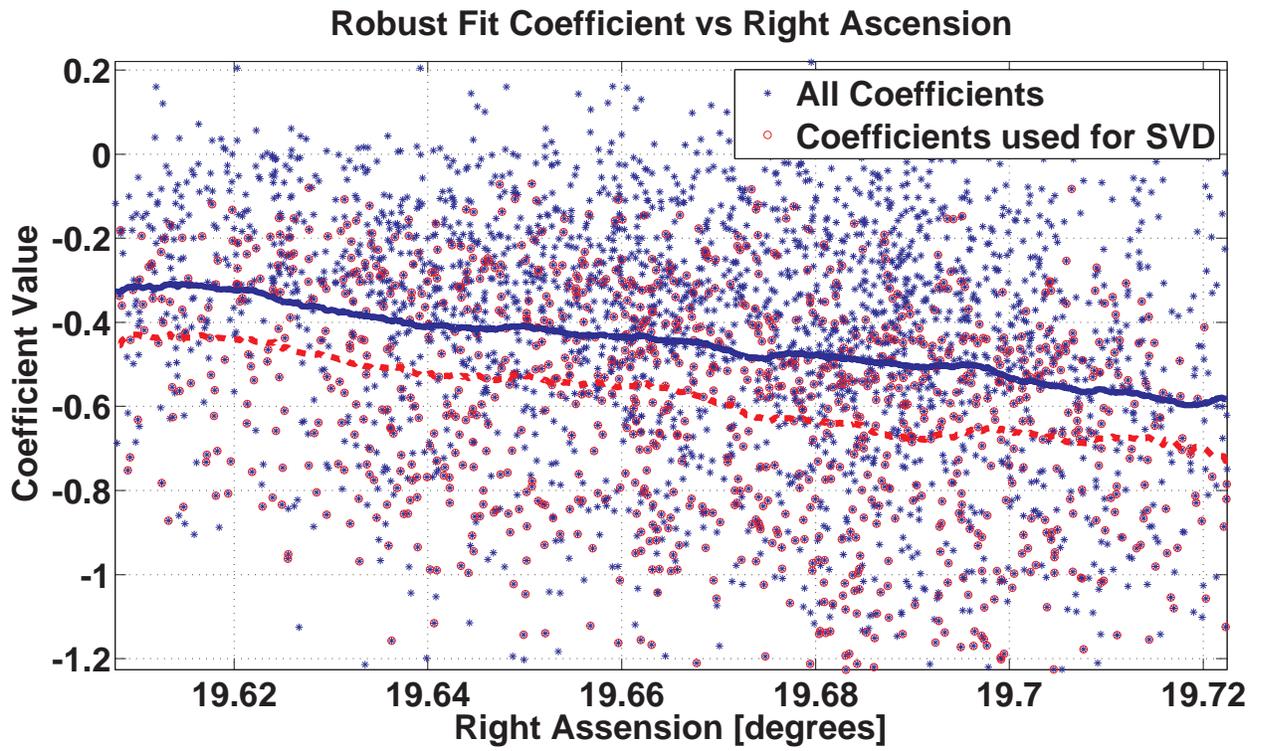}
\caption{Robust fit coefficients for Basis Vector 1 for all targets (Blue stars) and only those targets used for SVD
(Red circles) plotted against Right Ascension.}
\label{fig:coeff2VsRA}
\end{figure}

\begin{figure}
\epsscale{1.0}
\plotone{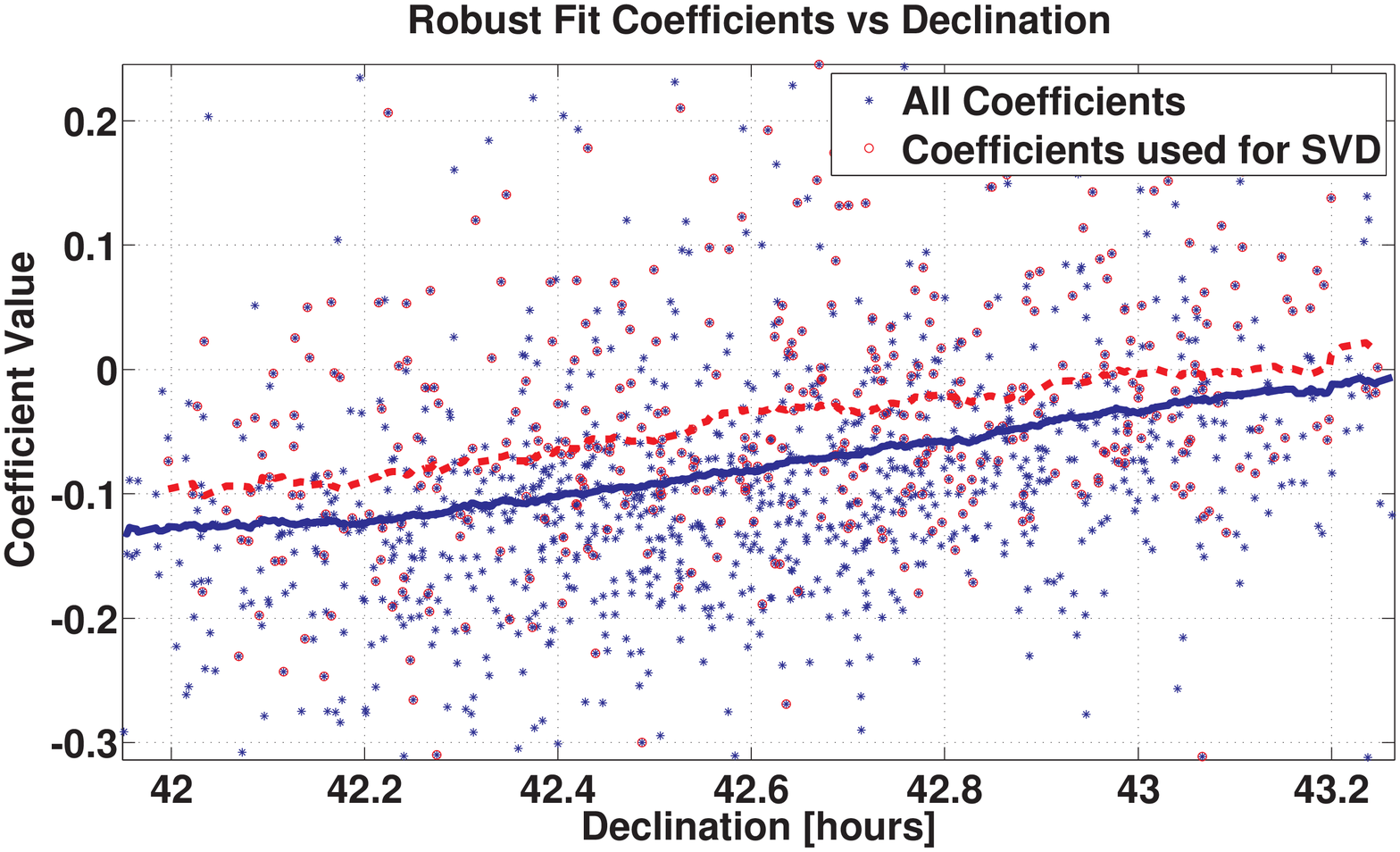}
\caption{Robust fit coefficients for Basis Vector 1 for all targets (Blue stars) and only those targets used for SVD
(Red circles) plotted against Declination.}
\label{fig:coeff2VsDec}
\end{figure}

Figures~\ref{fig:coeff2VsK_p}, \ref{fig:coeff2VsRA} and \ref{fig:coeff2VsDec} show the robust fit coefficients
for a basis vector plotted against $K_{p}$, RA and Dec. The blue star data is for all targets whereas the red
circle data is just for those targets remaining for SVD after the cuts discussed in
section~\ref{s:findingBasisVectors}. The solid blue and dashed red curves in Figures~\ref{fig:coeff2VsRA} and
\ref{fig:coeff2VsDec} are the travelling window means of the blue star and red circle data respectively.  The
cuts clearly produce a bimodal distribution in $K_{p}$ for this basis vector. A simple Gaussian fit would not
reproduce this and demonstrates that the systematic trends are correlated with $K_{p}$ but variable targets
are masking the true correlation when a simple robust fit is performed. The correlations in RA and Dec are
also evident but to a lesser extent.  Notice also that the mean (solid blue curves) are biased compared to the
dashed red.  This is again because the variable targets are masking the true trends in the data. 

Some basis vectors exhibit stronger trends in $K_{p}$, RA or Dec but not necessarily
all three simultaneously as is expected if the different systematics represented by the basis vectors
have different instrumental sources.  Plotting different Basis Vectors and/or channels reveals different trends
and correlations.

We want to mainly rely on targets in the neighborhood around the target we are fitting, referred to as the 
\emph{Target Under Study} (TUS), in RA, Dec and
$K_{p}$ space when generating the prior PDF. If we simply found an evenly
weighted PDF then a large cluster of targets with a certain coefficient value, even if non-local to the TUS,
would always dominate the peak of the prior PDF. We therefore use a weighted probability density estimate
based on the Standardized Euclidean Distance between targets $\xvector$ and $\yvector$,
\begin{equation}
D = \sqrt{\left(\xvector - \yvector\right) \boldsymbol{\Lambda}^{-1} \left(\xvector - \yvector\right)^{\mathrm{T}}}
\label{eq:distanceMetric}
\end{equation}
where $\boldsymbol{\Lambda}$ is a diagonal matrix whose diagonal elements give the relative weighting for each
dimension.
A straight normalization in each dimension by its standard deviation would result in equal weighting of all
three dimensions, but we wish to overemphasize the prior PDF in dimensions that exhibit greater correlations, and
in our case, the robust fit coefficients exhibit a stronger correlation in $K_{p}$ than in RA or Dec.
The $\boldsymbol{\Lambda}$ matrix diagonals are therefore 
\begin{equation}
\Lambda_{i} = \frac{\mathrm{mad}\left(\thetavector_{i}\right)}{ S_{i}}
\end{equation}
where $\mathrm{mad}\left(\thetavector_{i}\right)$ is the median absolute deviation of the coefficient
distribution along dimension $i$
and $S_{i}$ is the scaling factor for dimension $i$, 
$S_{i} = \{ 2, \text{if } i \Rightarrow K_{p}  \text{\;or\;} 1, \text{otherwise.}\}$
The above weighting results in the $K_{p}$ dimension weighted twice as much as RA and Dec when generating the prior PDF.
That is, targets further away in the $K_{p}$ dimension are weighted proportionately less than in RA and Dec.
This effectively results in taking a tighter cut in $K_{p}$ space to emphasize the greater correlation in that
dimension. Since the PDF is weighted by this distance metric, the PDF will emphasize the correlation in $K_{p}$
and yet still be sensitive to the trends in RA and Dec.
The median absolute deviation is used instead of the standard deviation in order to ignore outliers.

\begin{figure}
\epsscale{1.0}
\plotone{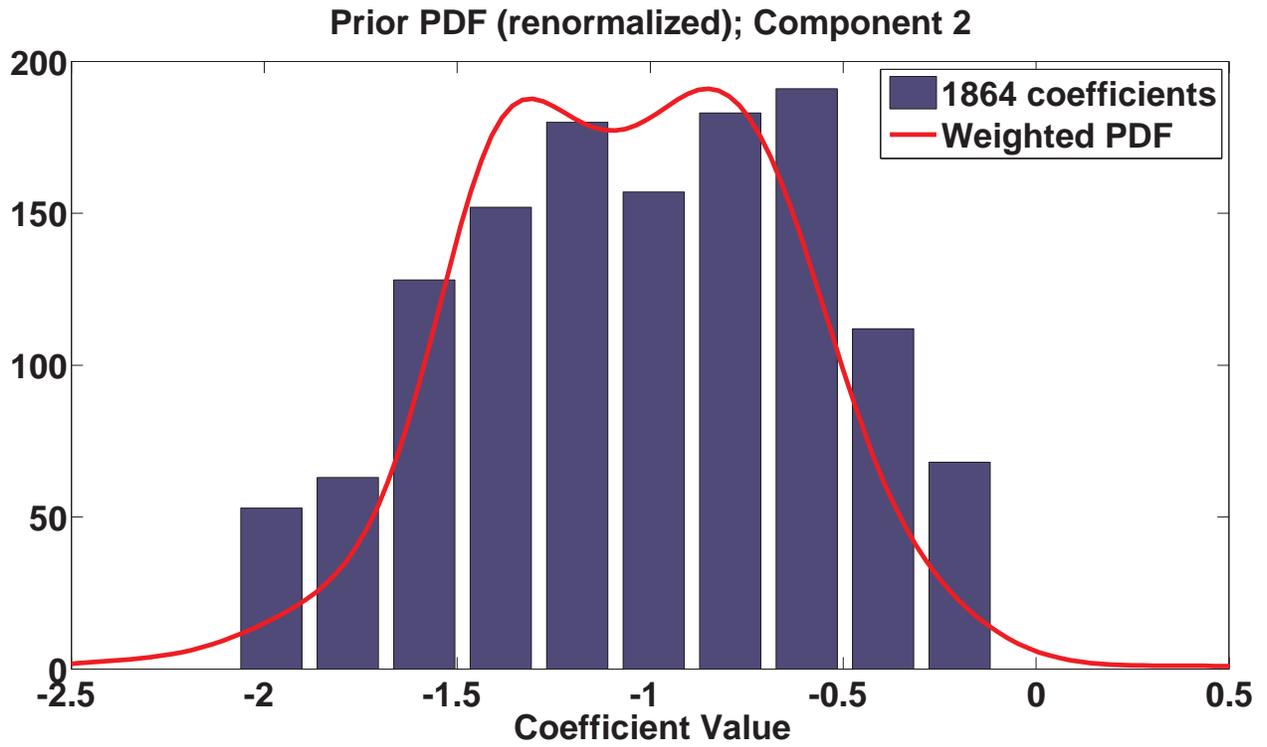}
\caption{Histogram of basis vector 2 robust fit coefficients for all 1864 targets on channel 2.1 and the weighted probability 
density for a particular
target. The weighting by distance in $K_{p}$, RA and Dec clearly affects the PDF.}
\label{fig:nonGaussPdf1}
\end{figure}

\begin{figure}
\epsscale{1.0}
\plotone{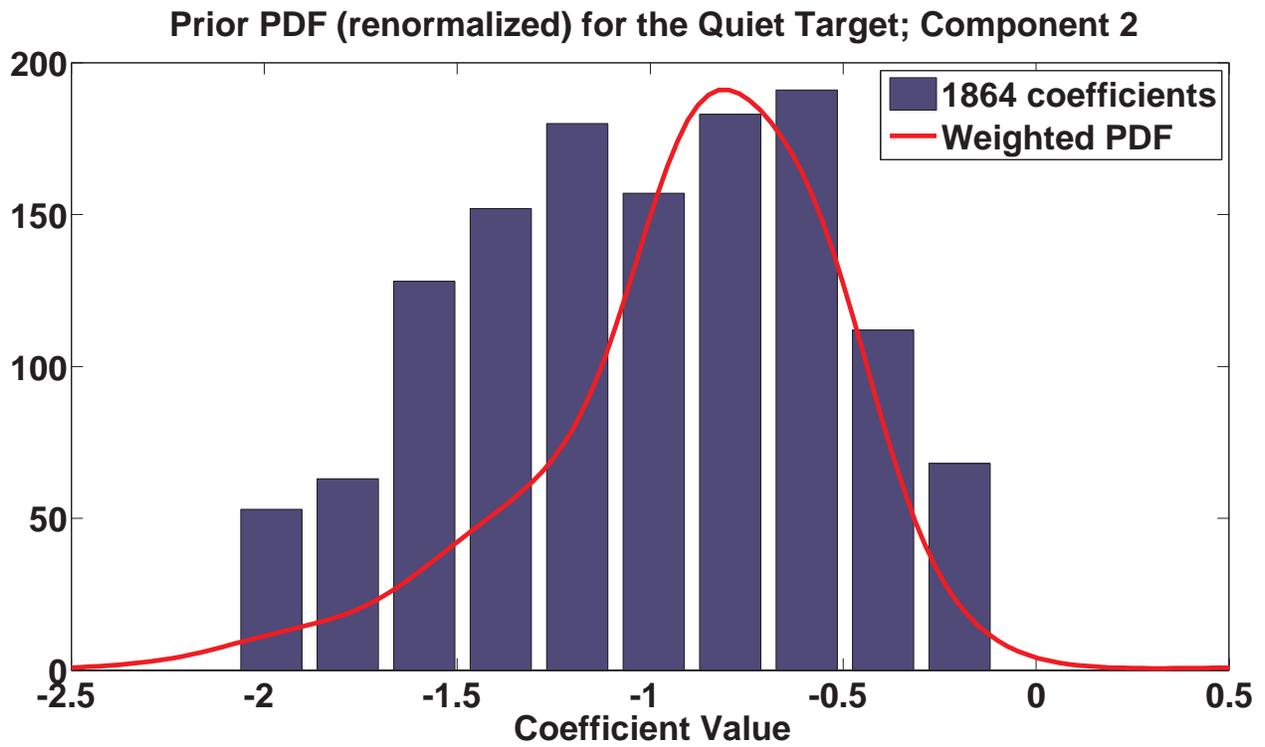}
\caption{Histogram of basis vector 2 robust fit coefficients for all 1864 targets on channel 2.1 and the weighted probability 
density for the Quiet Target shown in Figure~\ref{fig:veryQuietStar}.}
\label{fig:nonGaussPdfQuiet}
\end{figure}

\begin{figure}
\epsscale{1.0}
\plotone{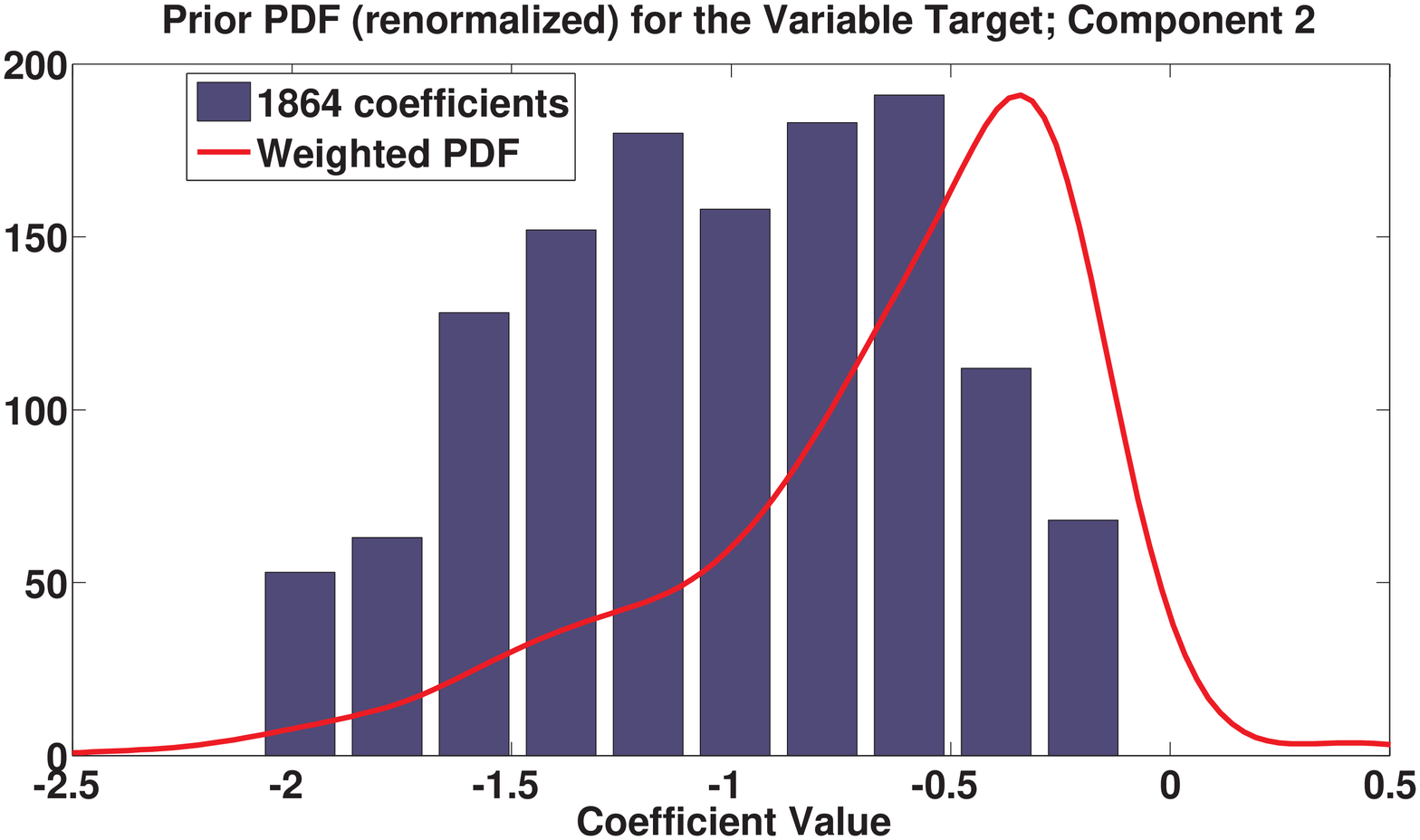}
\caption{Histogram of basis vector 2 robust fit coefficients for all 1864 targets on channel 2.1 and the weighted probability 
density for the Variable Target light curve shown in Figure~\ref{fig:veryNoisyStar}.}
\label{fig:nonGaussPdfVariable}
\end{figure}

The weighting by equation~\ref{eq:distanceMetric} and how it affects the prior PDF is illustrated in
Figures~\ref{fig:nonGaussPdf1}, \ref{fig:nonGaussPdfQuiet} and \ref{fig:nonGaussPdfVariable}. The later two
being the prior PDFs for the same two targets in Figures~\ref{fig:veryQuietStar} and \ref{fig:veryNoisyStar}.
The blue histogram in all three figures is exactly the same since they are generated from the same
distribution of coefficients. However, the prior PDF (red curve) is dramatically different. In
Figure~\ref{fig:nonGaussPdf1} a bimodal PDF is evident due to targets nearby to the TUS containing two
clusters, around -1.3 and -0.85, and suggests that the coefficient value for the TUS should be one of these
two values. Which value that is actually chosen will be dependent on the form of the conditional PDF and the
weighting of the prior PDF as discussed in Section~\ref{ss:findingWeighting}. 
In Figure~\ref{fig:nonGaussPdfVariable} the targets near the TUS have coefficients
clustered around -0.34 which is far from the peak in the unweighted PDF. Using the unweighted PDF would have
completely missed the actual systematic trend in the data near the TUS.  The log of the prior PDF is plotted in
these figures for direct comparison with equation~\ref{eq:mapEmpiricalSoln} which results in a compression of
the PDF near the top. 

In summary, the prior PDF is developed by generating a 3 dimensional weighted distribution of robust LS fit
coefficients in RA, Dec and $K_{p}$ space. This methodology makes no assumptions on the form of the PDF,
Gaussian or otherwise, and allows PDC-MAP to identify and characterize the form of the systematic trends
across the full distribution of targets.

\subsection{Finding the Weighting Parameter $\mathbf{W_{pr}}$}\label{ss:findingWeighting}

\begin{figure}
\epsscale{1.0}
\plotone{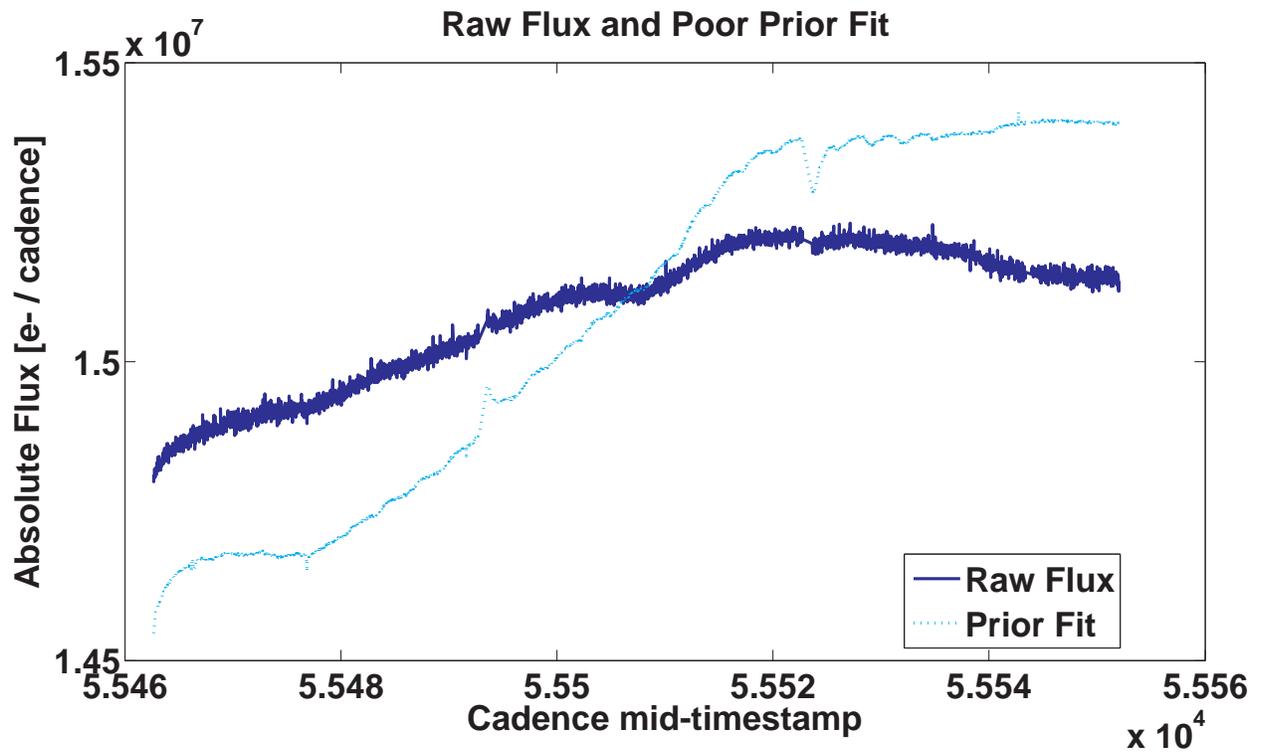}
\caption{An example of a poor prior PDF fit to the trend in the target.}
\label{fig:badPrior}
\end{figure}

For each light curve the weighting parameter, $\mathbf{W_{pr}}$, in equation~\ref{eq:mapEmpiricalSoln} is
an empirical weighting parameter that is principally based on the variability of each target. The greater the
variability the greater we need to constrain the least-squares fit. However, there is another
complication. A fit to the prior PDF is not always a good fit to the trend in the target. The reason for this
disagreement is currently being investigated. One factor influencing the ``goodness'' of the prior fit is the
sparseness of the targets in certain regions in Ra, Dec and $K_{p}$ space. A sparse distribution will result
in poor prior statistics. There are also other unknown causes resulting in poor priors for some targets and so 
an additional parameter in
the prior weighting is an evaluation of the ``goodness'' of the prior fit. The goodness is evaluated using a method
similar to the variability calculation above. The prior fit is compared to a 3rd order polynomial fit to
the light curve with a soft-wall cutoff using the following equation
\begin{equation}
G_{\text{pr}} = \begin{cases} 1 - \left(\frac{G_{\text{raw}}}{\alpha_{G}}\right)^{3},
                    &\text{if } G_{\text{raw}} < \alpha_{G} \\ 0, &\text{otherwise}\end{cases}
\label{eq:priorGoodness}
\end{equation}
where $G_{\text{raw}}$ is the ``raw'' goodness given by
\begin{equation}
G_{\text{raw}} = \text{std} \left( \frac{\left(F_{\text{pr}} - F_{\text{poly}}\right)}
     {\text{mad} \left( y - F_{\text{poly}} \right) } - 1 \right),
\label{eq:rawPriorGoodness}
\end{equation}
and $F_{\text{pr}}$ and $F_{\text{poly}}$ are the prior PDF fit and the 3rd order polynomial fit to the
data respectively. Normalization by the median absolute deviation (mad) of the polynomial fit removed light curve allows for
a comparison of the difference between the polynomial fit and the prior fit with respect to the variance of
the target. The soft cutoff is to ensure that small changes in the light curve will not have dramatic changes
in the weighting. The scaling parameter $\alpha_{G}$ is determined by when the deviation
of the prior fit to the polynomial fit
becomes too poor 
to be useful in constraining the Posterior fit. An example of a poor Prior Fit is given in
Figure~\ref{fig:badPrior}. Notice how both the long term trend and the Earth-point recoveries 
are much larger in the prior fit than in the
actual data. Examples
such as this are in the minority, but frequent enough to require the additional test for prior goodness.
It could be proposed that this target is trending downward cancelling out the upward trend of the prior fit.
This is unfortunately not the case. Examination of Kepler Season Quarters 6 and 8 reveal that this target is not
experiencing a general trend in Quarter 7. The prior fit is indeed poor and should not be used to any large
degree.
The unfortunate side effect of the prior goodness test is that \emph{PDC-MAP is less sensitive to very long term trends in
the data}. A true long term trend in the data that cancels out the systematic trend can confuse the prior
goodness metric which would interpret the fit as a bad prior. The only way to surely know the actual long
term trend is to examine multi-quarter data. Future versions of the PDC module may indeed provide this functionality.

The resultant full form to the prior weighting is
\begin{equation}
\mathbf{W_{\text{pr}}} = V^{\beta_{V}} * G_{\text{pr}}^{\beta_{G}}.
\label{eq:priorWeighting}
\end{equation}
The parameters $\beta_{V}$ and $\beta_{G}$ being scaling factors for the Variability and Prior Goodness
respectively. Future work includes fully characterizing the conditions where the prior fit is poor and thereby
remove the unfortunate need for an empirical prior ``goodness'' metric.

In cases where the prior goodness is near zero the fit reverts to a reduced robust fit where the number of
basis vectors is limited to just the first several (default is 4). A MAP fit has the pleasant feature where a
large number of basis vectors can be used. The prior PDF restricts the fit from drifting drastically in
function space searching the large set of basis vectors for a combination that reduces the bulk RMS at the
expense of distorting stellar features.  If the prior cannot be used then there is no such restriction and the
posterior PDF becomes a least-squares fit, so a more limited number of basis vectors must be used in order to
constrain the fit.  The first several basis vectors have very strong trends in most of the data and have low
noise components so they are generally safe to use even with an unrestricted least-squares fit. It is also
generally true that a target with a bad prior is so because the target is quiet and any small deviation in the
prior from a true trend is very noticeable and the prior is neither necessary or desirable to use.

If the target is below the variability threshold shown in Figure~\ref{fig:variabilityScatter} then the target is
very quiet and in many cases the use of the prior fit only worsens the fit over a least-squares fit and so the
prior weighting is zeroed.  This is due to the prior fit never being an exact match to the target trend and even
small deviations can ``pull'' the posterior fit away from a good fit.  In such cases there is little risk of a
quiet target biasing a least-squares fit away from a proper cotrending fit.
The majority of targets do not fall into either of the above two cases and the prior is used to the
degree dictated by the prior weight, and a Bayesian MAP fit is performed.

\subsection{Maximization of the Posterior PDF}

Once the prior weighting is determined the posterior PDF can be assembled using
equation~\ref{eq:mapEmpiricalSoln}. Due to the empirical, and therefore non-analytical, form of the prior PDF
the posterior must be maximized numerically. In general, a multidimensional maximization is difficult and time
consuming due to the risk of only finding local maxima. Fortunately, due to the use of SVD, the basis vectors are
all orthogonal so the various coefficients $\hat{\theta}_{i}$ can be maximized sequentially. The process is
therefore straightforward. The strongest singular vector is maximized first and then all
subsequent singular vectors are maximized in turn.

\begin{figure}
\epsscale{1.0}
\plotone{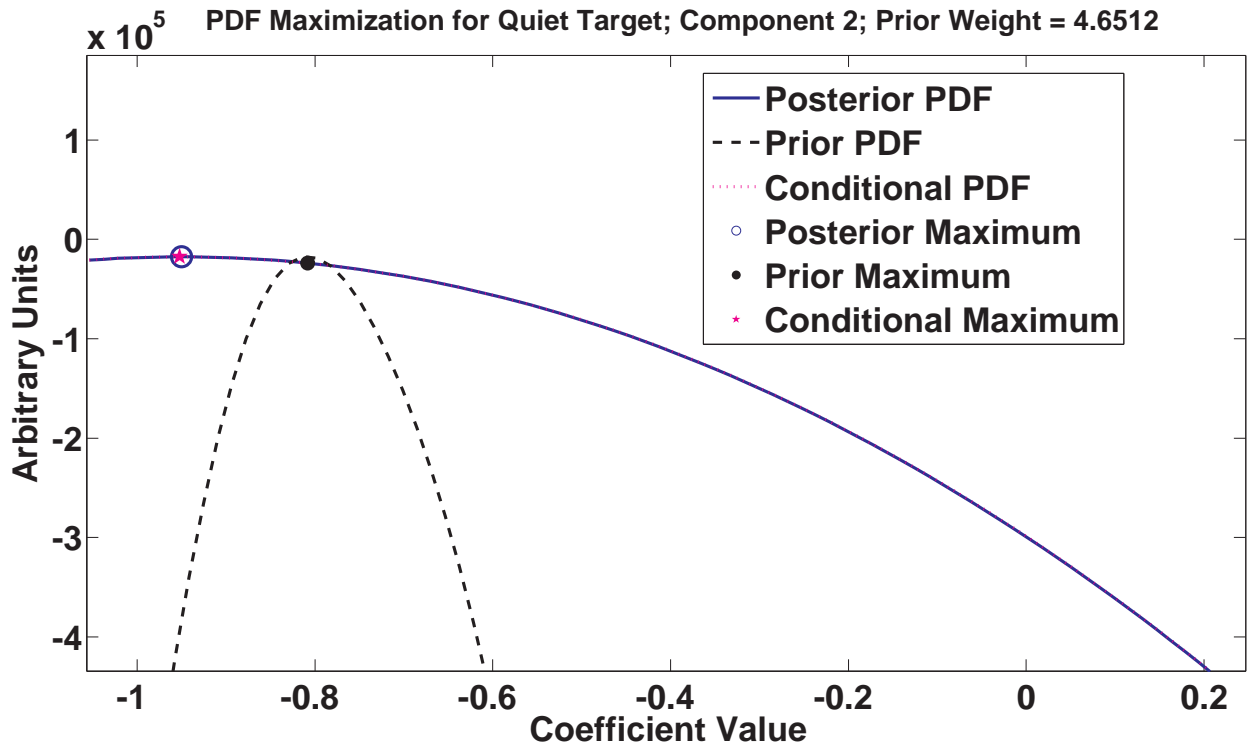}
\caption{Posterior, prior and conditional PDFs for the Quiet Target. The prior and conditional curves have been 
renormalized to the same scale as the Posterior for legibility. This target is quiet so the prior PDF
weighting is low and does
not influence the posterior by much. The maximum of the posterior is therefore very close to the
conditional maximum. The width of the prior PDF can also influence its height and amount of influence on the conditional.}
\label{fig:posteriorQuiet}
\end{figure}

\begin{figure}
\epsscale{1.0}
\plotone{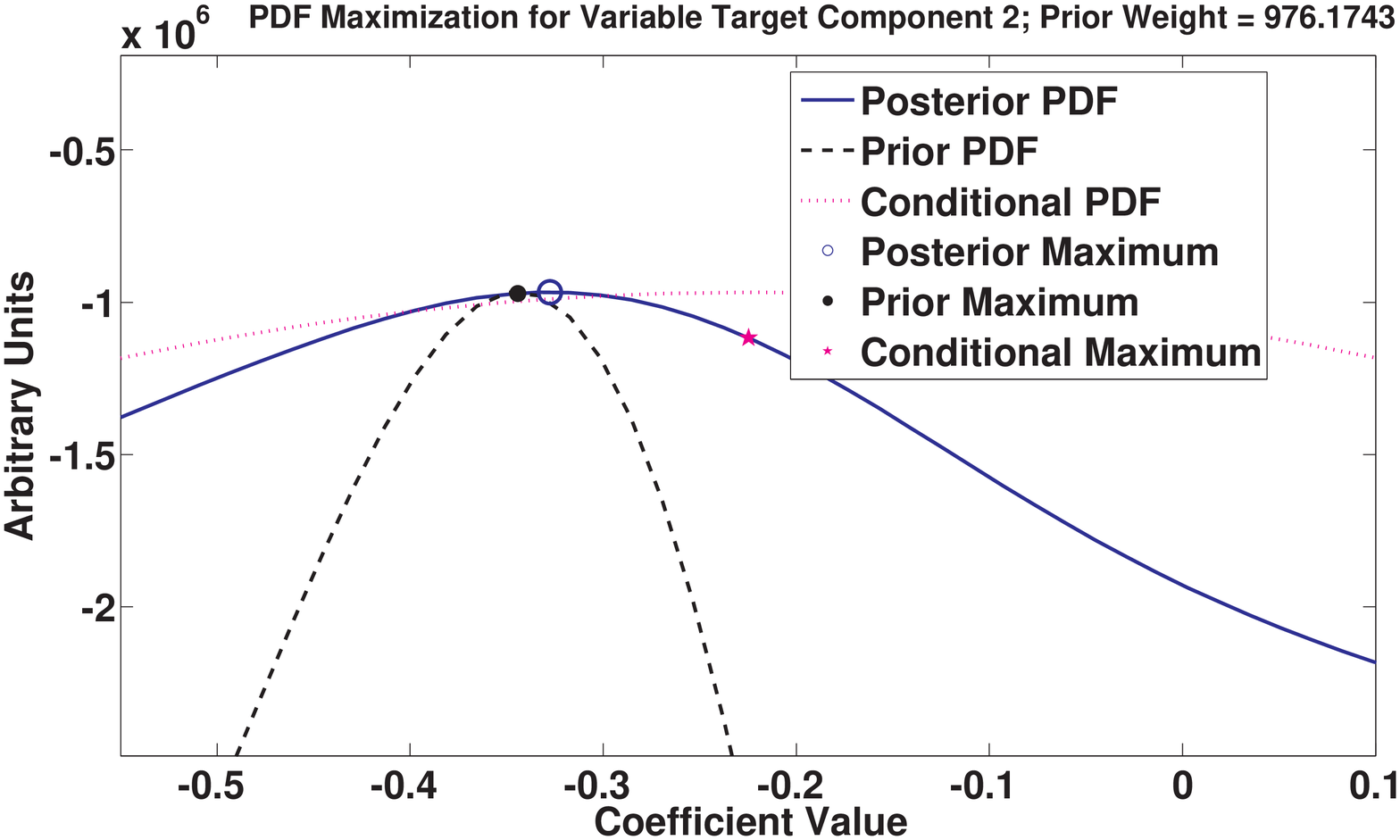}
\caption{Posterior, prior and conditional PDFs for the Variable Target. The prior and conditional curves have been 
renormalized to the same scale as the posterior for legibility. This target is highly variable so the prior
PDFs is highly weighted at 976 and it influences the posterior considerably. 
The maximum of the posterior is therefore close to the
prior maximum.}
\label{fig:posteriorVariable}
\end{figure}

Following along with the same two examples of a quiet and a variable target, Figures~\ref{fig:posteriorQuiet}
and \ref{fig:posteriorVariable} show the final posterior PDF along with the prior and conditional PDFs. 
The black dots and magenta stars are the maxima of
the prior and conditional PDFs and the blue circle is the maximum of the posterior PDF. Due to
the varying scales of the three curves the prior and conditional curves have been renormalized to the same
scale as the posterior for illustration purposes. The conditional fit has the smooth quadratic form
characteristic of a least-squares fit. The prior appears Gaussian on this scale but in general is not. The
title of each plot gives the prior weight. For the quiet target the variability is low so the prior weight is
only 4.65 resulting in a minor correction to the conditional curve and so the conditional and posterior curves
are virtually identical. Also, in cases where the width of the Prior PDF is large the maximum is low and it makes little
contribution to the posterior. A wide prior is equivalent to saying that the prior information results in little
added information to a good fit. The extreme case being a flat prior PDF which provides no additional
information.
This is in contrast to the variable target
where the conditional maximum is near the Prior PDF, meaning the fit almost completely relies on the prior
data, due to the high weight on the Prior PDF of 976, resulting in
only a slight influence of the conditional fit.
In the case of the variable target in particular, if the prior PDF was not determined using a weighted
distribution (in $K_{p}$, RA and Dec space) then the maximum of the prior would have been at about 0.6 as shown
by the unweighted histogram in Figure~\ref{fig:nonGaussPdfVariable}. This would have resulted in a poor fit to the
systematics. The actual prior fit takes into account the location of the Target Under Study and the systematic trends in
targets nearby to the TUS. 

\begin{figure}
\epsscale{1.0}
\plotone{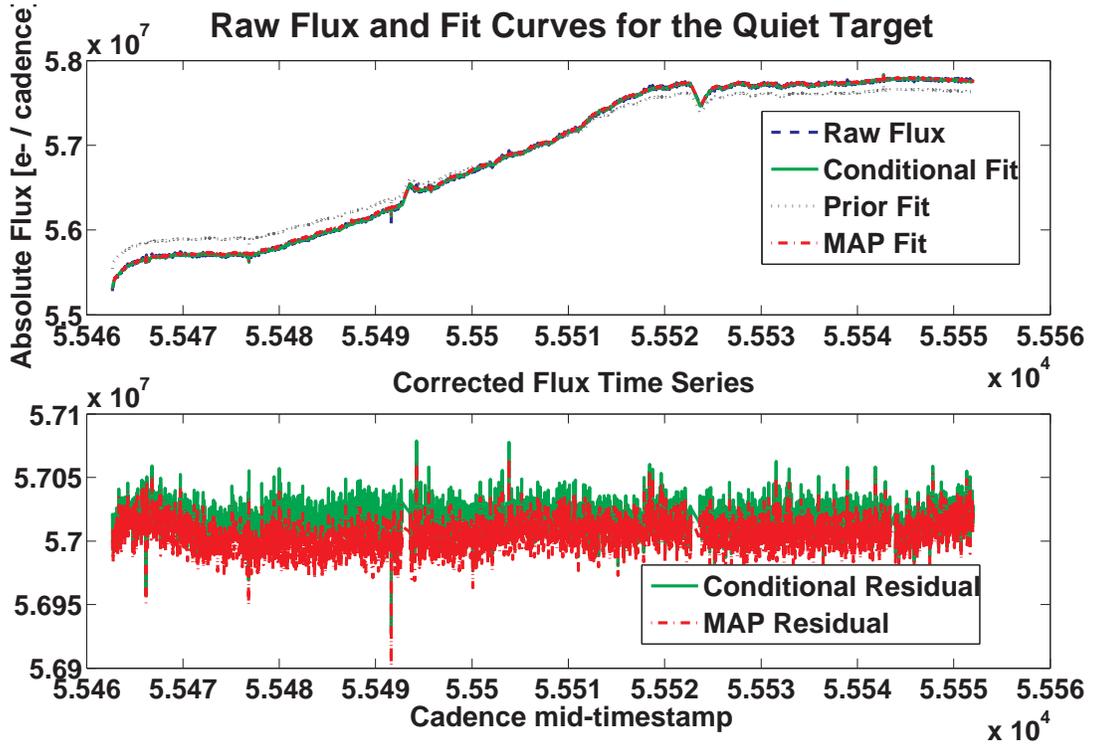}
\caption{Resultant fits to the Prior, conditional and posterior (MAP) PDFs for the Quiet Target. Target
variability is low at 1.17 and $W_{pr}$ is also low at 4.65. Quiet targets
rely principally on the conditional PDF.}
\label{fig:fitQuiet}
\end{figure}

\begin{figure}
\epsscale{1.0}
\plotone{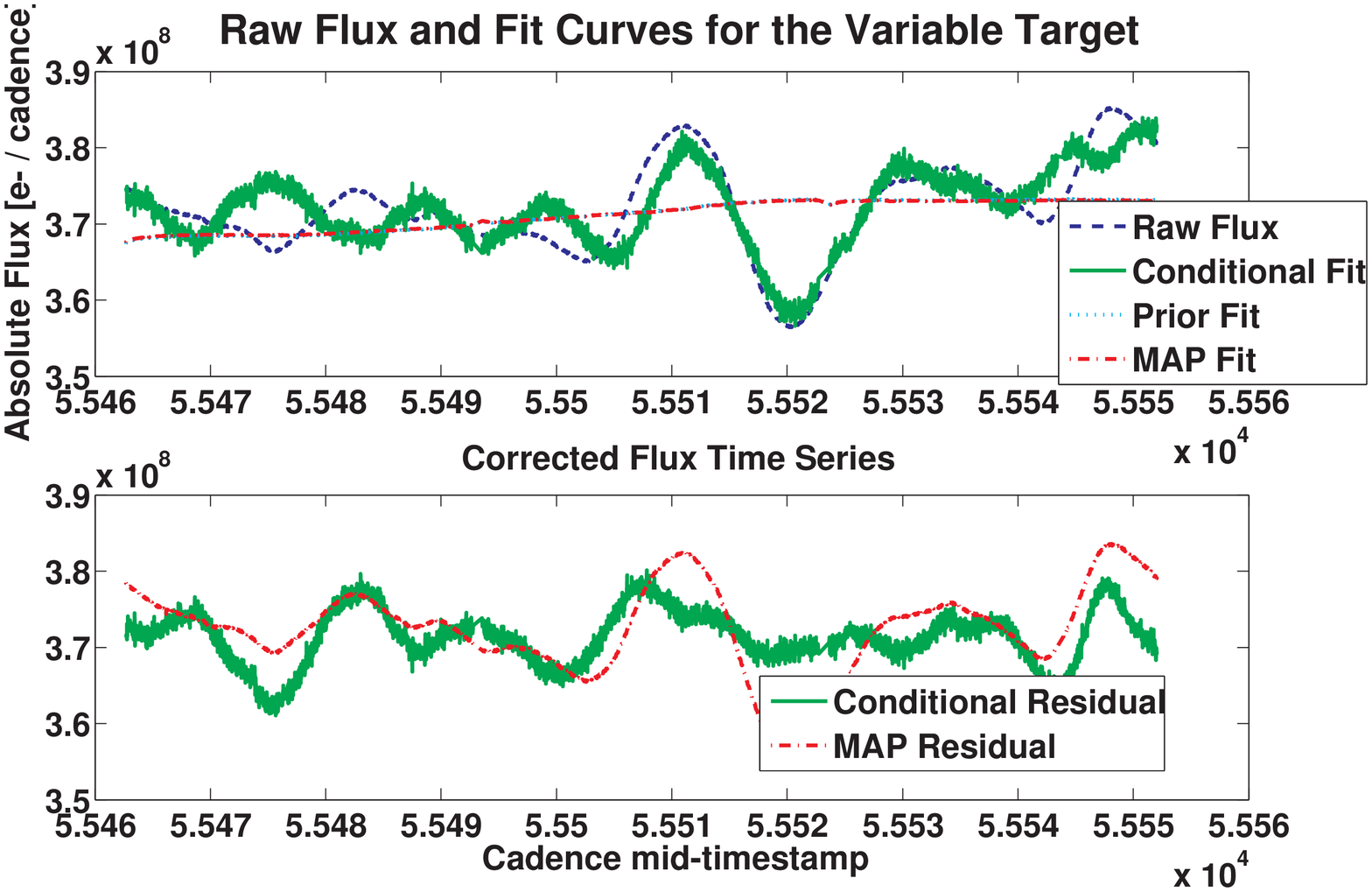}
\caption{Resultant fits to the prior, conditional and posterior (MAP) PDFs for the Variable Target. Target
variability is high at 30.25 and $W_{pr}$ is also high at 976.2. Variable targets
rely principally on the prior PDF. It is evident that the Posterior (MAP) fit finds the systematic trend yet
preserves the variability.}
\label{fig:fitVariable}
\end{figure}

Once the maximum of the posterior PDF is found for each basis vector the MAP fit is a linear
combination of the basis vectors. The resultant fits are in Figure~\ref{fig:fitQuiet} and
\ref{fig:fitVariable}. For the Quiet Target all three fits roughly overlap the actual trend in the data.  The
prior fit is not an exact match and the slight disagreement is to be expected since the prior is purely formulated using
targets other than the TUS. For a quiet target such as this one highly weighting the prior PDF would result in
a degradation of the fit and so instead the PDF relies mainly on the conditional PDF (i.e. the red dashed and
green solid curves overlap). The resultant light curve after the trend removal is in the bottom figure for both the
MAP fit and the conditional fit, the later being a least-squares fit. For the quiet target notice how the
resultant curve is near featureless above the noise floor.  Some slight artifacts are not fully removed and
methods to correct these are discussed in Section~\ref{s:futureImprovements}.  In the case
of the variable target the conditional PDF results in a fit that attempts to remove all features in the light
curve, whereas the Prior PDF correctly identifies just the systematic trends in the data. In this case it is
beneficial to rely principally on the prior PDF. The prior cannot be well discerned in the figure because it
lies under the MAP fit.  The
conditional fit also introduces a considerable amount of noise into the corrected light curve due to it being
constructed more from the lesser basis vectors (shown in Figure~\ref{fig:basisVectors}) which contain larger
noise components.

\subsection{Propagation of Uncertainties}

Propagation of uncertainties is not necessarily straightforward because a
covariance matrix is difficult to formulate for an empirical prior PDF. 
As a first approximation the propagation can be assumed to be
through a least-squares solution -- which is close to the solution for most targets.
If $C_{\text{raw}}$ and $C_{\text{cot}}$ denote the covariance matrices 
for the temporal samples of
the raw and cotrended flux time series for a given target, then the uncertainties may be propagated
(disregarding the uncertainty in the mean level which can be considered to be negligible) by
\begin{equation}
C_{\text{cot}} = T_{\text{cot}} C_{\text{raw}} T_{\text{cot}}^{T},
\end{equation}
where the transformation $T_{\text{cot}}$ is defined by
\begin{equation}
T_{\text{cot}} = \left( I - H H^{T} \right).
\end{equation}
$H$ being the same design matrix as in equation~\ref{eq:mapEmpiricalSoln}.
This is overly conservative since the posterior PDF is more constrained than a simple least-squares fit.
A more accurate propagation of uncertainties would take into account the attenuation of the uncertainties due
to the prior PDF. 

\section{Future Improvements}\label{s:futureImprovements}

The algorithm as presented works phenomenally well
for the majority of light curves in the Kepler FOV. Overall PDC performance is discussed in a companion
paper~\citep{stumpe}. However,
problems do arise. Remediations to these problems are discussed in this section.

One of the main issues with the current method is different types of systematic effects are represented in
each basis vector shown in Figure~\ref{fig:basisVectors}. For example, the long term trends associated with
DVA and seasonal changes should not be represented by the same basis vectors as heater cycles and Earth point
thermal recoveries. Given that these different systematics behave on different time-scales a reasonable
solution is to band-split the light curves and generate separate basis vectors for each band. This method is
currently in development.

The current method relies solely on Singular Value Decomposition to generate the basis vectors after cuts on
quiet and correlated targets.  SVD is a reliable and often used tool to generate basis vectors that describe
highly correlated trends in data. However, it does have its detractions. For non-Gaussian systematic trends,
SVD is not ideal. Methods such as Independent Component Analysis as described in~\citep{waldmann} can
potentially better de-convolve independent systematic sources. 

Occasionally a single target can dominate one or more basis vectors. The resulting basis
vectors essentially contain all stellar variability and noise of the offending target making the basis vector
impotent in removing systematics. The entropy cleaning step removes the offending targets, but the source of
the problem is partially a result of the normalization by the median value of each light
curve. For the small number of dim targets with appreciable noise, the noise is overly represented in the
basis vectors. Normalization by the standard deviation or noise floor of the light curves would limit the
problem of over-represented targets, however at the expense of not equally normalizing the light curves by flux
intensity which can introduce other problems.

We have also discovered that over the channel field of view groups of targets exhibit similar systematics that are
distinct from other targets so specific clusters of targets can be identified with similar trends.
Using the same basis vectors for all targets is not ideal in this situation. We are investigating using a
Hierarchical Clustering method such as described in~\citep{jain} to isolate the clusters and develop basis
vectors separately for each cluster.

The prior PDF generation is based on correlations in Stellar Magnitude, Right Ascension and Declination but in
some targets the prior is poor. It is
to be expected that systematics are also correlated with other stellar and instrumental parameters.  A full
parametric study finding these correlations is to be performed to identify hidden variables that further characterize the
systematics. One strongly suspect hidden variable is sub-pixel centroid motion. 

Kepler collects both long cadence data at $\sim$30 minute intervals and short cadence data at $\sim$60 second intervals.
No more than 512 short cadence targets are collected at any time and are spread over the entire field of view so the
number of short cadence targets per channel is small and at most about a dozen. A dozen is
too small of a sample for the prior PDF to be properly formulated. We are investigating ways to extend PDC-MAP to
short cadence data. Options include using the prior PDF from long cadence data and using a single reference
ensemble drawn from all 512 targets.

\section{Conclusions}
PDC-MAP dramatically improves Kepler's ability to understand the
properties of parent stars. It preserves stellar signals and minimizes the noise while removing the systematic
errors
that can mask transit signals.
PDC-MAP therefore ultimately improves Kepler's primary mission of detecting Earth-like
planets. But PDC-MAP also improves the Kepler data's utility to the broader astrophysical community.
Non-planet finding studies such as asteroseismology is greatly benefited by PDC-MAP's ability to preserve
stellar signals in the light curves.


\acknowledgments

Funding for this Discovery Mission is provided by NASA's Science Mission Directorate.  We thank the thousands
of people whose efforts made \emph{Kepler's} grand voyage of discovery possible. The Kepler Science Office and
Science Working Group work diligently in analyzing data products and have provided great insight into data
reduction methods and data processing.  We especially want to thank the \emph{Kepler} SOC staff who helped
design, build, and operate the \emph{Kepler} Science Pipeline for putting their hearts into this endeavor.
The authors would also like to thank the very productive conversations we have had with the Suzanne Aigrain group at
Oxford University, especially Stephen Reece, Stephen Roberts and Amy McQuillan.

{\it Facilities:} \facility{Kepler}.

\bibliographystyle{apj}

\end{document}